\documentclass{article}
\usepackage[margin=2.5cm]{geometry}
\usepackage{graphicx} 
\usepackage[export]{adjustbox}
\usepackage{amsmath,amssymb}
\usepackage{bm}
\usepackage{tikz}
\usepackage{subcaption}
\usepackage{hyperref}
\usepackage{lipsum}
\usepackage{cleveref}
\usepackage{subcaption}
\usepackage{float}
\usepackage[most]{tcolorbox} 

\newtcolorbox[auto counter]{optionalnote}[2][]{
    parbox=false,
    colbacktitle= white,
    colback=green!5!white,
    colframe=white!45!black,
    coltitle=black,
    enhanced,
    attach boxed title to top left={yshift=-1mm},
    title={\thetcbcounter.~#2}
,#1}

\newtcolorbox{highlight-result}[1][]{
 parbox=false,
 boxrule=0pt,top=0pt,bottom=0pt,
colback=blue!7!white,
enhanced,#1}

\tcbset{colback=blue!7!white, colframe=white!50!white,top=0mm,bottom=0mm,right=0mm,left=0mm}

\title{Designing Band Gaps with Randomly Distributed Sub-Wavelength Helmholtz Resonators}
\author{Paulo S. Piva$^1$, Art L. Gower$^1$, I. David Abrahams$^2$}
\date{%
    $^1$School of Mechanical, Aerospace and Civil Engineering, University of Sheffield, Mappin Street, Sheffield, England, S1 4DT Sheffield, United Kingdom\\%
    $^2$Department of Applied Mathematics and Theoretical Physics, University of Cambridge, Wilberforce Road, Cambridge, England, CB3 0WA, United Kingdom\\[2ex]%
    \today}

\begin{document}

\maketitle

\begin{abstract}
    It is well-known that band gaps, in the frequency domain, can be achieved by using periodic metamaterials. However it has been challenging to design materials with broad band gaps or that have multiple overlapping band gaps. For periodic materials this difficulty arises because many different length scales would have to be repeated periodically within the same structure to have multiple overlapping band gaps. Here we present an alternative: to design band gaps with disordered materials. We show how to tailor band gaps by choosing any combination of Helmholtz resonators that are positioned randomly within a host acoustic medium. One key result is that we are able to reach simple formulae for the effective material properties, which work over a broad frequency range, and can therefore be used to rapidly design tailored metamaterials. We show that these formulae are robust by comparing them with high-fidelity Monte Carlo simulations over randomly positioned resonant scatterers.
    
\end{abstract}

\section{Introduction}
\label{sec:intro}

Metamaterials are artificially fabricated composite materials which have properties not found in nature; they are designed for specific and targeted applications~\cite{Metamats_review_caoustic}. Wave motion control is an important application for electromagnetics, elastodynamics, and acoustics. There are many examples of fabricated materials that can distort or bend the path that waves propagate along, allow negative refraction~\cite{Fancy_science_paper}, block specific frequencies~\cite{book_metamats_waves}, or cloak (i.e.\ render invisible) certain parts of space~\cite{Cummer_2007,PARNELL_TOM_CLOAK,Fancy_science_paper_2,NORRIS_hidding_stuff}. In recent years, the focus on elastic and acoustic metamaterial theory and design has broadened to include non-reciprocal wave propagation~\cite{Fancy_nature_paper}, acoustic lenses~\cite{Numerics_lens}, and optimal wave-absorbing layered media~\cite{Optimal_abs}.  

In many applications, the design of metamaterials for wave control is often a heuristic process, which encourages heavy optimisation and machine learning techniques in the literature \cite{Kumar2020,Mao2023}. However, even after long and extensive computations to find an optimal metamaterial arrangement, small manufacturing changes/defects can lead to errors that significantly alter the band structure. For this reason, these materials require elaborate manufacturing techniques \cite{manufacturing,fan2021review}. 



One possible strategy to achieve multiple band gaps is to combine different types of sub-wavelength resonators, each with a different resonance frequency. However, to date there has been no clear strategy or simple formula for the effective properties of a mix of different types of resonators. It has only been possible to derive the effective properties of a periodic array of identical sub-wavelength resonators \cite{Krynkin_2011,Dave's-latticeI,Dave's-latticeII}.

In this paper, we present a disordered (random) metamaterial, which can be designed to have broad as well as multiple localised (overlapping) low-frequency band gaps, that are robust to small changes in its microstructure. We show how to design such metamaterials and, importantly, derive simple explicit formulae for the band gaps which do not require heavy optimisation. We achieve this by combining ensemble averaging techniques~\cite{Effective_density,Gower_2021,gower_smith_parnell_abrahams_2018} together with asymptotic homogenisation. 

In disordered materials, it is more straightforward to add a mix of resonators, with the simplest case being just a uniform random mix of the resonators. In terms of fabrication, the main difficult is to make the resonators themselves and not to precisely place them. The drawback of disordered materials is that their scattering response from any one configuration of these resonators is complicated. In this paper, we greatly simplify this complication by taking an ensemble average: that is, we calculate the average wave by averaging over all possible positions and properties of the resonators. As a result, a band gap in this context is where the average wave transmission is approximately zero, but there may still be some transmission in the form of a completely speckled wave \cite{sheng2007introduction}.  




%

\begin{figure}[!ht]
    \centering
    \begin{subfigure}[t]{0.42\textwidth}
        \centering
        \includegraphics[width=\linewidth]{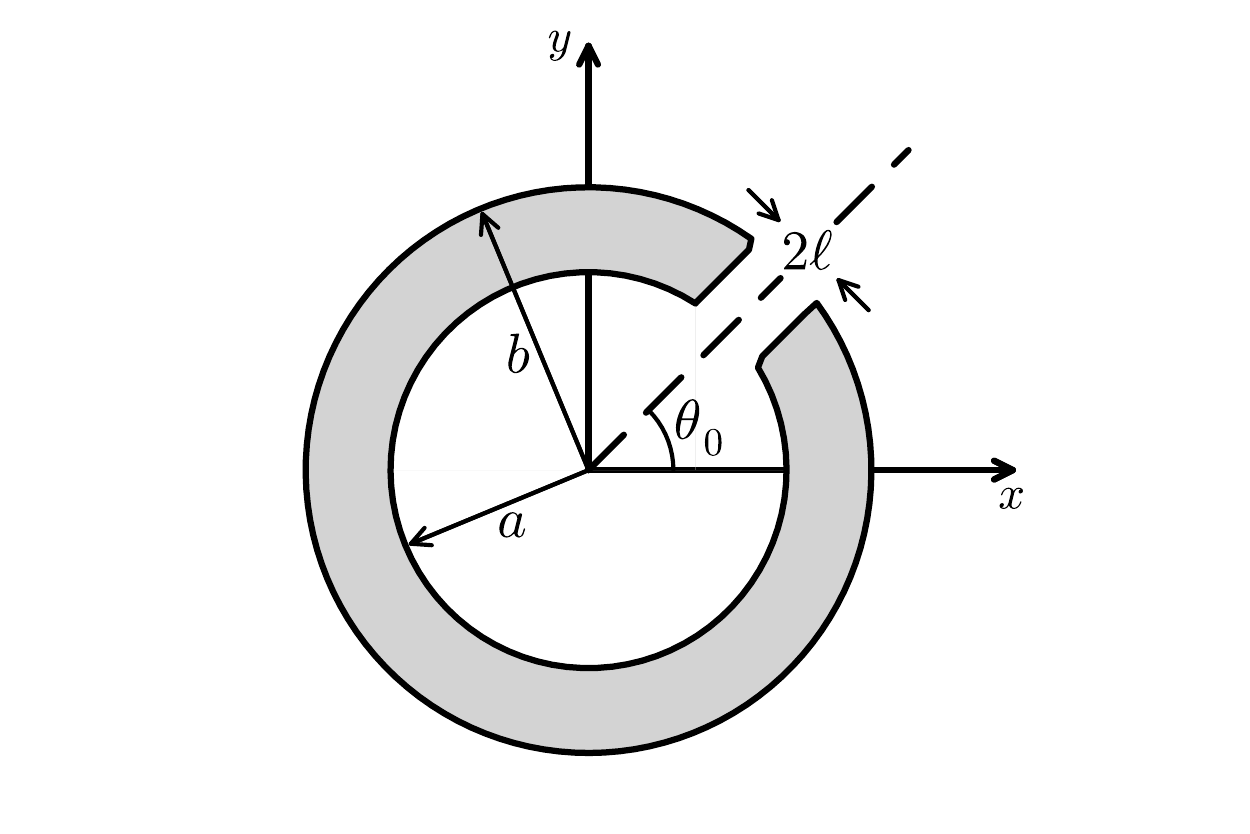}
    \end{subfigure}%
    \hspace{1.0cm}
    \begin{subfigure}[t]{0.5\textwidth}
        \centering
        \includegraphics[width=\linewidth]{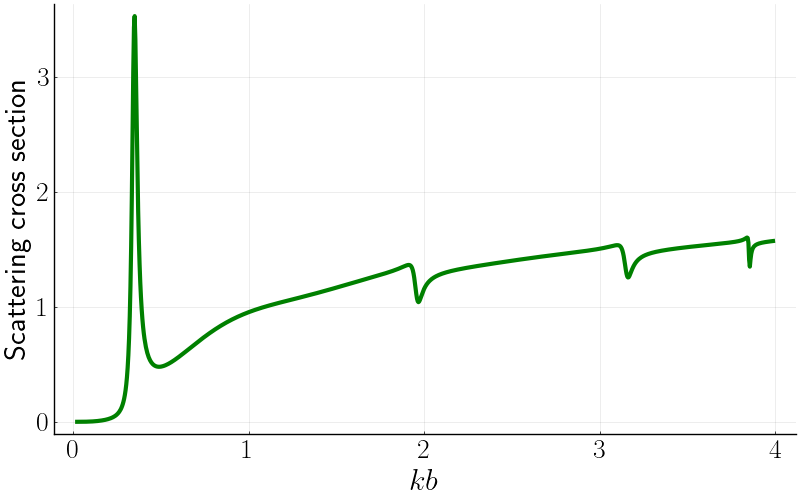}
    \end{subfigure}
    \vspace{-1cm}

    (a) \hspace{5.8cm} (b)  \hspace{6.8cm}
    \vspace{0.1cm}
    \caption{The left figure (a) shows an illustration of a typical split-ring resonator. The right figure (b) shows the scattering cross section of a sound-hard 2D split-ring resonator with an aperture size of $2\ell = 0.1b$, and $k$ is the wavenumber of the
background medium.}
    \label{fig:split-ring_resonator}
\end{figure}

In this paper, we use as an example a split-ring resonator, shown in \Cref{fig:split-ring_resonator}a, because its scattered wave has been deduced from first principles \cite{Atom-Helmholtz, Split_ring_lipmann_scwinger, Krynkin_2011}. It is a type of Helmholtz resonator, characterised by a cavity-neck system with resonance peak located in the sub-wavelength range ($kb \leq 1$), shown in \Cref{fig:split-ring_resonator}b. An example of using split-ring resonators in a periodic lattice is deduced in \cite{Atom-Helmholtz,Dave's-latticeI,Dave's-latticeII}.

%

To reach simple formulae, which are needed to easily design materials, asymptotic homogenisation is essential \cite{Krynkin_2011,homogenisation_book, parnell_abrahams_2008, Salvatore_Ganacci,Effective_props_coated_cylinders_in_fluid}. It is typically used to enforce that the resonators are sub-wavelength, as we also do in this work. However, care must be taken to accurately capture the resonance.


%
\begin{figure}[ht!]
    \centering
    \begin{subfigure}[t]{0.45\textwidth}
        \centering
        \includegraphics[width=\linewidth]{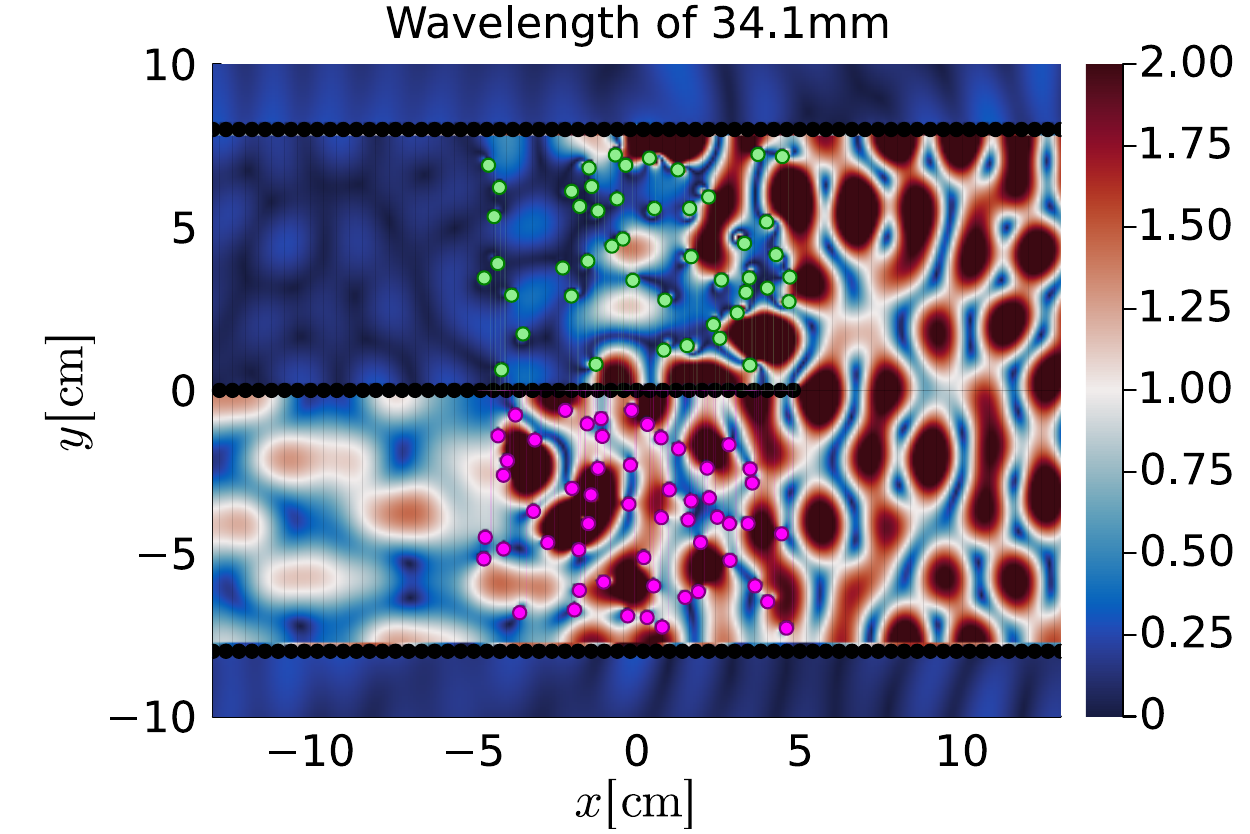}
    \end{subfigure}%
    ~ 
    \begin{subfigure}[t]{0.45\textwidth}
        \centering
        \includegraphics[width=\linewidth]{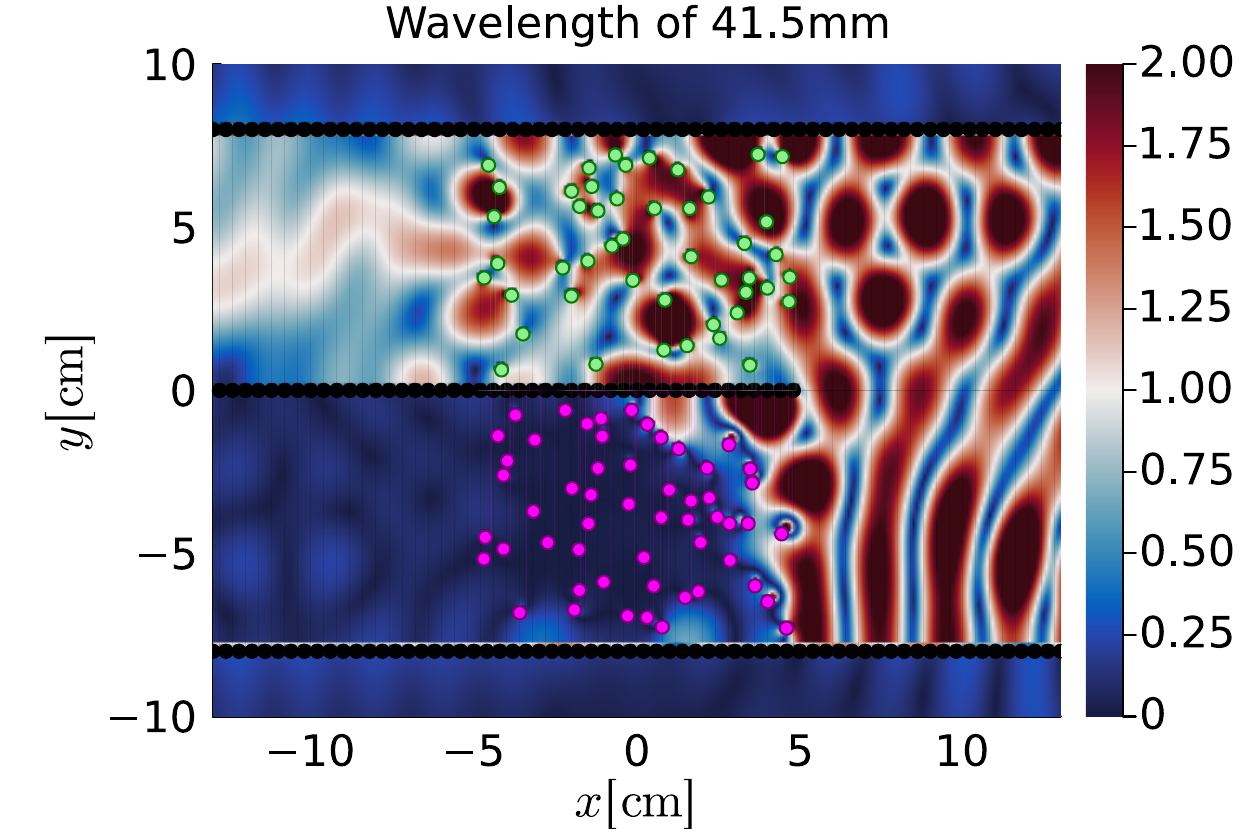}
    \end{subfigure}
    \begin{subfigure}[b]{\textwidth}
        \centering
        \includegraphics[width=0.8\linewidth]{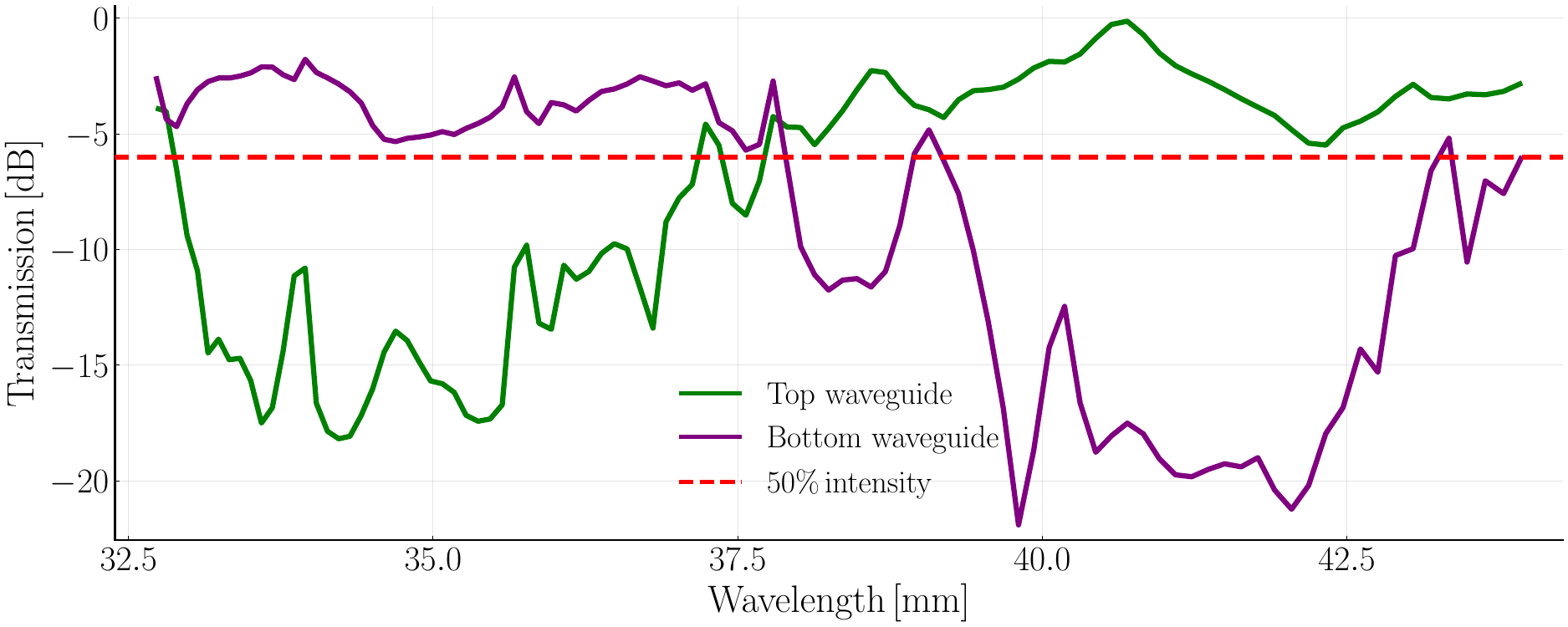}
    \end{subfigure}
    \caption{The top two images show a very simple frequency demultiplexer we designed using Helmholtz resonators to illustrate our results. In both images an incident planar wave (of amplitude 1) travels from right to left inside a waveguide which gets split into two waveguides at $x = 5$cm. The top waveguide filters out waves with wavelengths close to 34.1mm due to the scattering by resonators represented by green circles, with an aperture of $2\ell = 0.2$mm. Likewise, the bottom waveguide filters out waves with wavelengths close to 41.5mm due to the scattering by resonators represented by magenta circles, with an aperture of $2\ell = 0.04$mm. All resonators are thin-walled ($a=b=2$mm in \Cref{fig:split-ring_resonator}a) and solid, randomly placed and oriented in air at the entrance of each waveguide.
    The colour bar shown is the amplitude of the total wave (incident plus scattered).
    The graph below shows the transmitted intensity by each waveguide in dB. Wavelengths from 33 to 38mm are mostly transmitted through the lower waveguide, while wavelengths from 38 to 43mm are mostly transmitted through the upper waveguide.
    }
    \label{fig:all_mighty_frequency_splitter}
\end{figure}
%



To showcase the design process using our formulae, and motivate the reader for the rest of the paper, we present an acoustic version of a device that splits a signal into two different channels, or waveguides, dependent on its frequency components shown in \Cref{fig:all_mighty_frequency_splitter}. These devices are known as frequency demultiplexers. They are used in digital computing and telecommunication networks, and are often designed via tailored periodic structures, as in \cite{Metamats_elasticity_frequency_splitter}, or through black-box optimisation, as in \cite{Piggott2015}. In contrast, as illustrated in \Cref{fig:all_mighty_frequency_splitter}, we are able to produce with little effort a low insertion loss ($\lesssim$ 5dB) and high contrast ($\sim10-20$dB) demultiplexer using only randomly placed and oriented resonators, which has similar performance compared to the microchip in \cite[figure4b]{Piggott2015}. 

Designing devices, such as the demultiplexer in \Cref{fig:all_mighty_frequency_splitter}, with randomly oriented resonators makes them robust to manufacturing defects in terms of the positioning of each resonator. This is because the formulae for the effective-properties are agnostic to the position and orientation of the resonators. Alternatively, the performance could be further enhanced by optimising the position of the resonators. 
 



\section{Results}
\label{sec:design}

In this section, we present our main results: the effective-properties of the metamaterial. Later, in \Cref{subsec:low_freq}, we show how to derive these results. The focus below is on how to use the effective-properties, followed by examples, and to present high-fidelity Monte-Carlo validation.

\begin{figure}[ht!]
    \centering
    \includegraphics[width=0.56\linewidth]{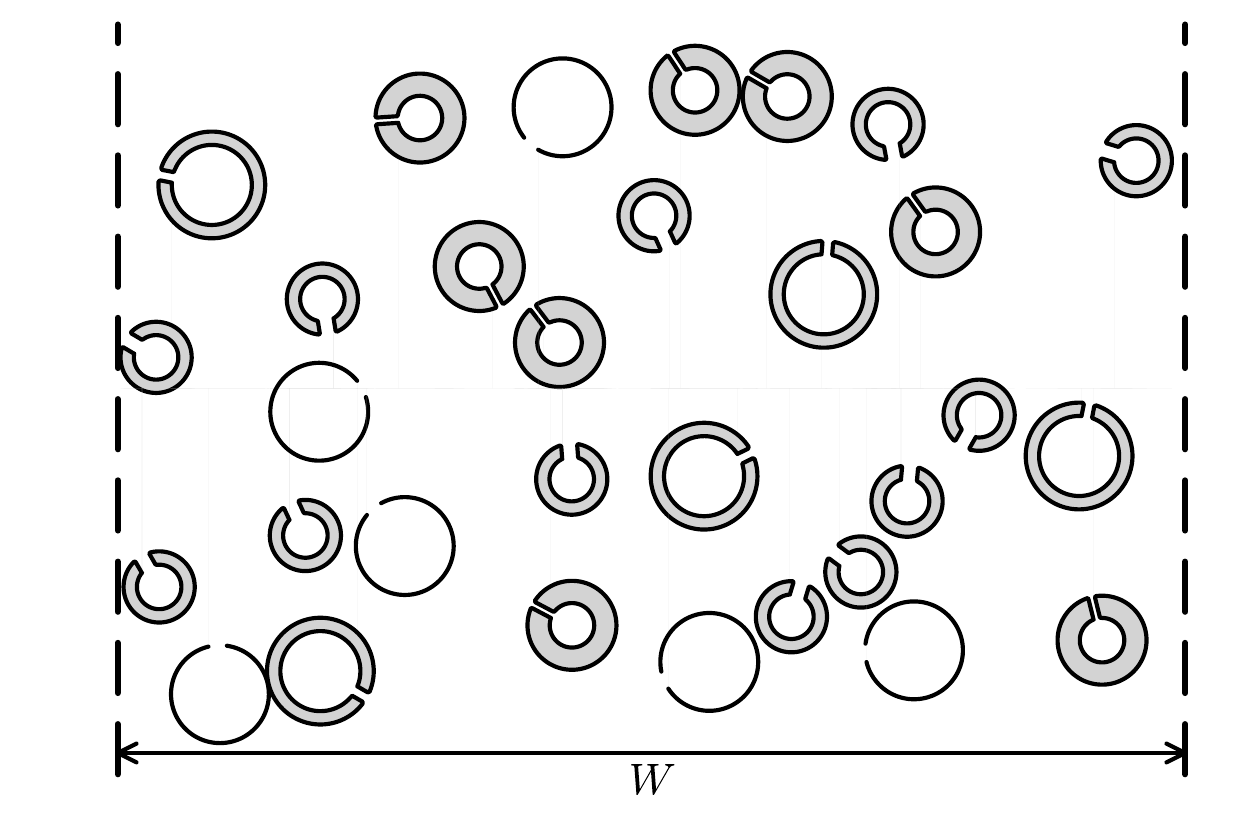}
    \caption{Representation of one possible configuration of split-ring resonators in a layer of width $W$. }
    \label{fig:meta-material}
\end{figure}

The resonators in the material can have varied sizes, geometry, and properties, although for ease of exposition we illustrate the case of circular sound-hard split-ring resonators in 2D, as shown in \Cref{fig:meta-material}. Let us consider $N$ different types of sound-hard Helmholtz resonators, which are randomly distributed in a homogeneous background medium (see \Cref{fig:meta-material}).
The properties of the resonators feed into the formulae for the effective bulk modulus ($\beta_\star$) and effective mass density ($\rho_\star$), which are given by:
\begin{equation}
    \label{eq:effective_formulae}
    \tcboxmath{
    \beta_\star(k) = \frac{\beta}{\left(1 - \varphi \right) + \sum_{j = 1}^{N} z(\lambda_j) \phi_j}
    }
\quad \text{and} \quad
    \tcboxmath{
    \rho_\star = \rho \frac{1 + \varphi}{1 - \varphi}
    },
\end{equation}
where $\beta$ and $\rho$ are the bulk modulus and mass density of the background medium, $\phi_j$ is the volume fraction of the $j$-th type of resonator\footnote{For two dimensions, the volume fraction reduces to the ratio of the area occupied by a phase to the total area of the material.}, with $\varphi = \sum_{j=1}^N \phi_j$ being the total volume fraction of the resonators (including their interiors). We call $z(\lambda_j)$ the \emph{resonance factor} of the $j$-th resonator type derived via careful matched asymptotic analysis \cite{Atom-Helmholtz, Dave's-latticeI}. The resonance factor depends only on the geometrical properties of the resonator and the wavenumber $k$ of the background medium. When hitting a resonance $z(\lambda_j)$ increases in magnitude, and when there is no contribution from the resonator then
$z(\lambda_j)$ is small. See \Cref{subsec:single_scattering} for more details and see \Cref{app:T-matrix} for a derivation of a formula for $z(\lambda_j)$ for split-ring resonators.

The effective bulk modulus $\beta_\star$ is frequency-dependent, and its behaviour depends on the geometry of each resonator in the mixture. $\beta_\star(k)$ is the most important parameter to design band gaps for metamaterials, since the effective mass density $\rho_\star$ does not depend on frequency and only increases when adding more resonators to the mixture. 

Below we show three examples on designing band gaps by using the formulae above. After the examples, we show how high-fidelity Monte-Carlo simulations closely agree with our simple formulae for a broad frequency range.

\subsection{Example band gap designs}
In this section we present three examples to show how to design band gaps using a layer of randomly placed resonators, as illustrated in \Cref{fig:meta-material}. For all the examples we assume that the background medium is air ($\rho =$1kg/m$^3$, $\beta=$117.6kPa).

\noindent \textbf{Example 1 -- changing volume fractions.} Consider a material filled with a single species of thin-walled split-ring resonator ($a = b$ and one fixed value of the aperture $2 \ell$ in \Cref{fig:split-ring_resonator}a). To help illustrate the results, we consider a layer of width $W$ (infinite length), shown in \Cref{fig:meta-material}, and calculate its effective transmission coefficient by assuming the material is homogeneous with the properties given by \eqref{eq:effective_formulae}. 

\begin{figure}[ht!]
    \centering
    \begin{subfigure}[t]{0.5\textwidth}
        \centering
        \includegraphics[width=\linewidth]{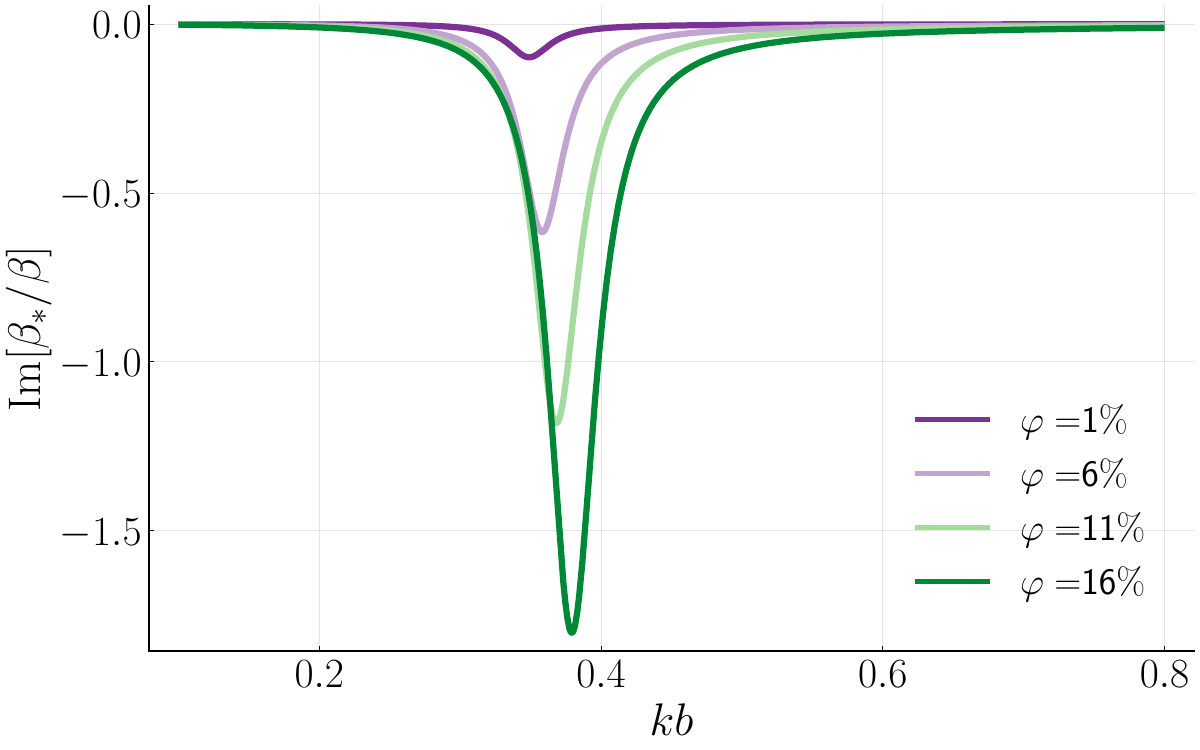}
    \end{subfigure}%
    ~ 
    \begin{subfigure}[t]{0.5\textwidth}
        \centering
        \includegraphics[width=\linewidth]{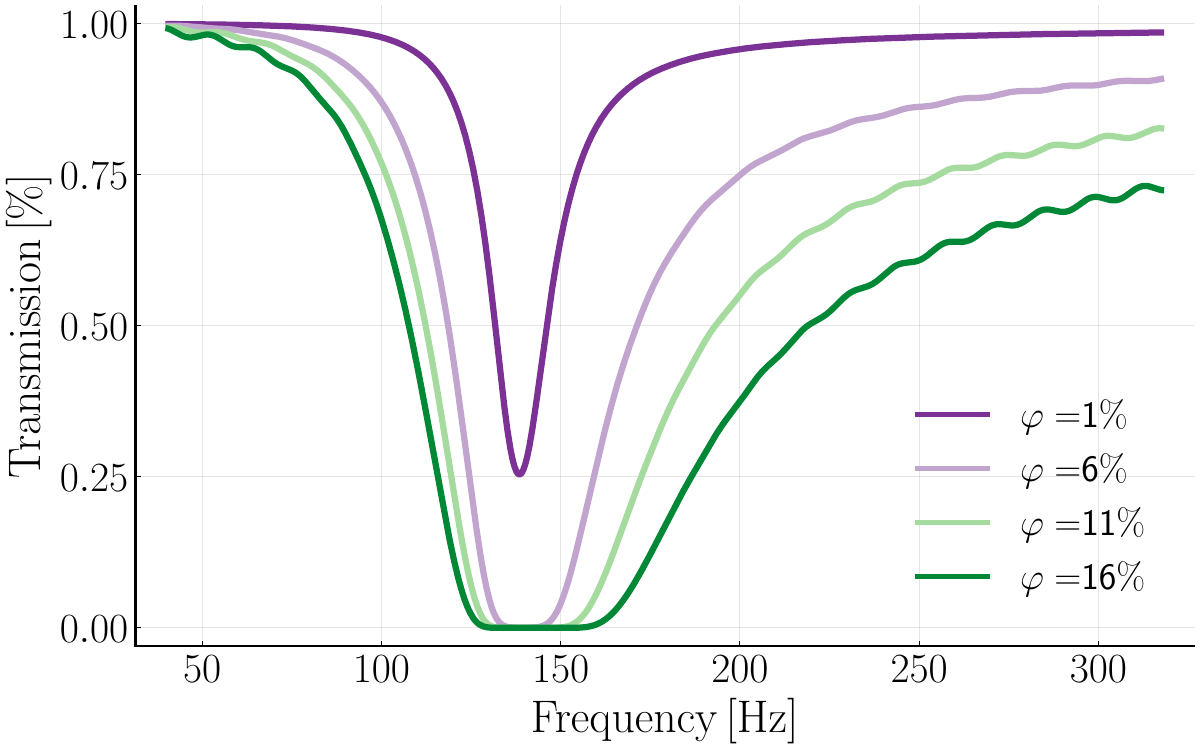}
    \end{subfigure}
    \vspace{-1.1cm}

    (a) \hspace{0.46\textwidth} (b) \hspace{0.4\textwidth}
    \caption{(a) shows the imaginary part of the effective bulk modulus $\beta_\star$ and (b) the magnitude of the transmission coefficient through a layer of width $W =$ 32mm. The layer is composed of thin-walled randomly-positioned Helmholtz resonators with radius $b =$ 0.4mm, and $k$ is the background wavenumber. The aperture size of the resonators is $\ell = 0.05b$. Each curves has a different volume fraction of resonators.}
    \label{fig:one-species-example}
\end{figure}

From the results,  shown in \Cref{fig:one-species-example}b, we see that only a small percent of resonators are needed to completely stop transmission through this layer. 
As the volume fraction increases, the minimum grows wider and shifts towards higher frequencies. With only $\varphi = 6\%$ volume fraction, we reach a band gap around 140Hz, which shows that a thin layer of the proposed metamaterial is a good candidate for a sound-insulating material. To broaden the band gap, it is possible to either increase the width of the metamaterial layer $W$, or the volume fraction of resonators $\varphi$ (see \Cref{fig:one-species-example}b).

Another lesson to learn from \Cref{fig:one-species-example}a is that the minimum of the curve $\varphi = 1\%$ is very close to the resonance frequency of a single thin-walled resonator, shown in \eqref{fig:split-ring_resonator}b. However, the minimum of $\mathrm{Im}[\beta_\star]$ the curves shifts to higher frequencies as $\varphi$ increases because of multiple scattering effects between the resonators. Being able to predict this minimum, with our formulae \eqref{eq:effective_formulae}, rather than finding it through trial and error, can save time and resources in most applications.

Finally, because \Cref{fig:one-species-example}a has non-dimensional independent variables, note that increasing the size of the resonators, while thickening their walls, would lead to the band gap shifting to a lower frequency range. 

\begin{figure}[ht!]
    \centering
    \begin{subfigure}[t]{0.5\textwidth}
        \centering
        \includegraphics[width=\linewidth]{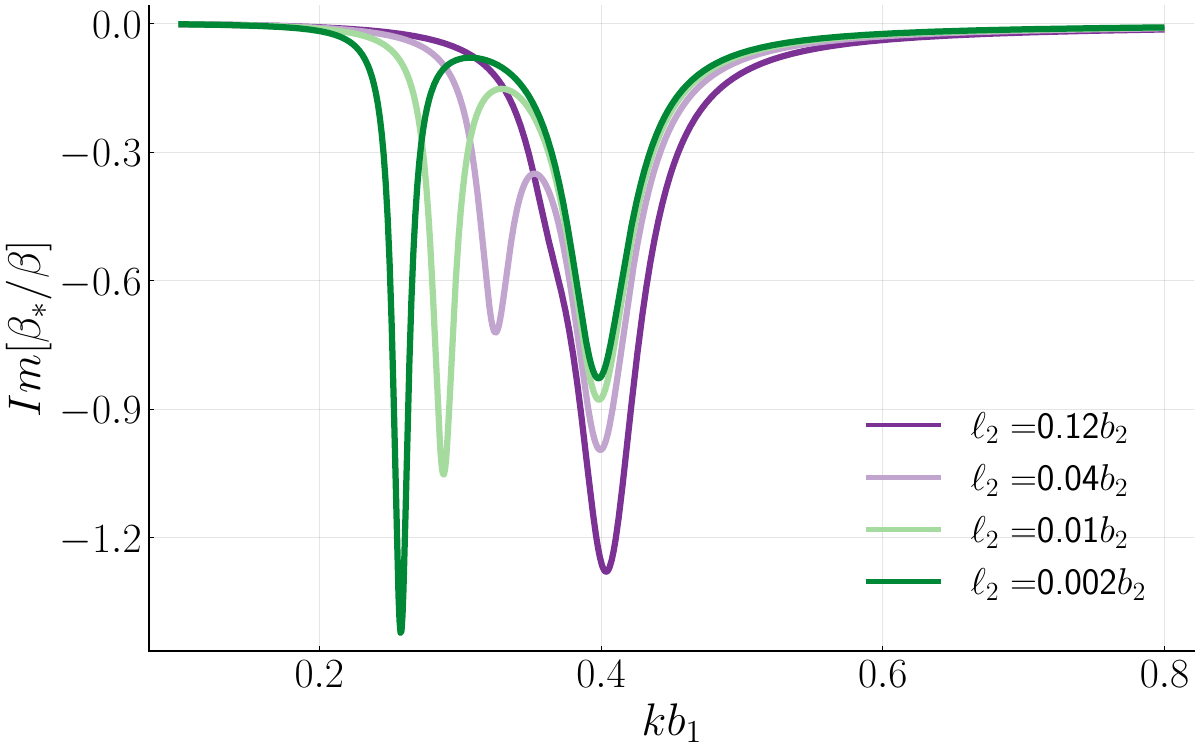}
    \end{subfigure}%
    ~ 
    \begin{subfigure}[t]{0.5\textwidth}
        \centering
        \includegraphics[width=\linewidth]{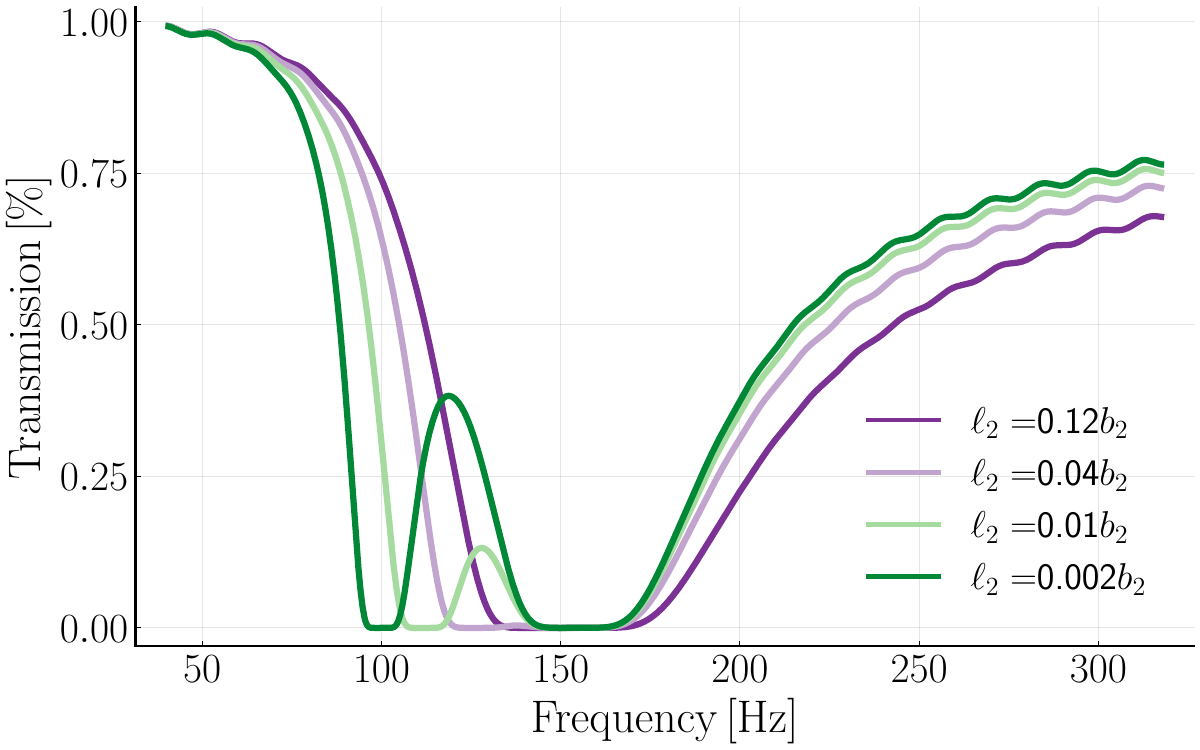}
    \end{subfigure}
     \vspace{-1.1cm}

    (a) \hspace{0.46\textwidth} (b) \hspace{0.4\textwidth}
    \caption{(a) shows the imaginary part of the effective bulk modulus and (b) shows the magnitude of the transmission coefficient through a layer of width $W =$ 32mm. The layer is composed of thin-walled random Helmholtz resonators with radius $b_1=$ 0.4mm and $b_2$ = 0.2mm. The total volume fraction of resonators is $\varphi = 15\%$, and half of them ($\phi_1 = 7.5\%$) have an aperture size of $\ell_1 = 0.1b_1$. Only the aperture size $\ell_2$ of the other half of the resonators changes between curves.}
    \label{fig:two-species-example}
\end{figure}

\noindent \textbf{Example 2 -- broadening band-gaps.} To achieve a broader band gap, or separate band gaps, we could use layers of random resonators of the type discussed in Example 1. Instead, we show a case with two types of thin-walled resonators to the formulae \eqref{eq:effective_formulae} with multiple types of resonators. The two types have different radii, $b_1 \neq b_2$, and aperture sizes, $\ell_1 \neq \ell_2$. The results are presented in \Cref{fig:two-species-example}.

In \Cref{fig:two-species-example}(a) we notice that when the aperture $\ell_2$ decreases, the graph of $\mathrm{Im}[\beta_\star]$ transitions from having one minima to two distinct minima. The appearance of the second dip is due to the resonance frequencies of the two types of resonators moving further apart as we decrease $\ell_2$. The overall result in the transmission (\Cref{fig:two-species-example}b) is a wide band gap, which separates into two thinner band gaps as $\ell_2$ decreases. 

\noindent \textbf{Example 3 -- shifting band-gaps.}  Here we consider a mix of three types of resonators, each with a volume fraction of $\phi_j = 4\%$ for $j=1,2,3$. We use three types just to show case that we can. Each resonator has a different aperture size, but the three types have the same outer and inner radius $b$ and $a$, for simplicity. Our goal is to show how the band-gaps shift when  the wall of the resonators become thicker. Increasing the thickness causes a non-intuitive change in the resonance which can only be predicted with our effective properties~\eqref{eq:effective_formulae}. 

\begin{figure}[ht!]
    \centering
    \begin{subfigure}[t]{0.5\textwidth}
        \centering
        \includegraphics[width=\linewidth]{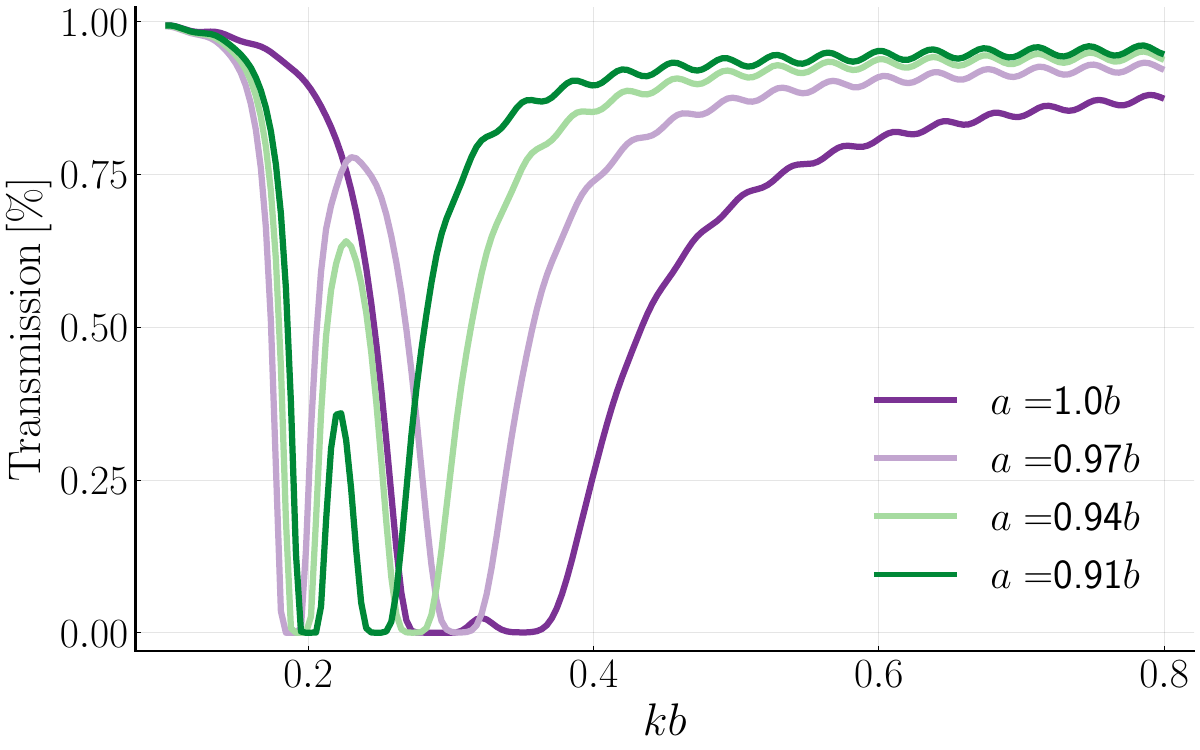}
    \end{subfigure}%
    ~ 
    \begin{subfigure}[t]{0.5\textwidth}
        \centering
        \includegraphics[width=\linewidth]{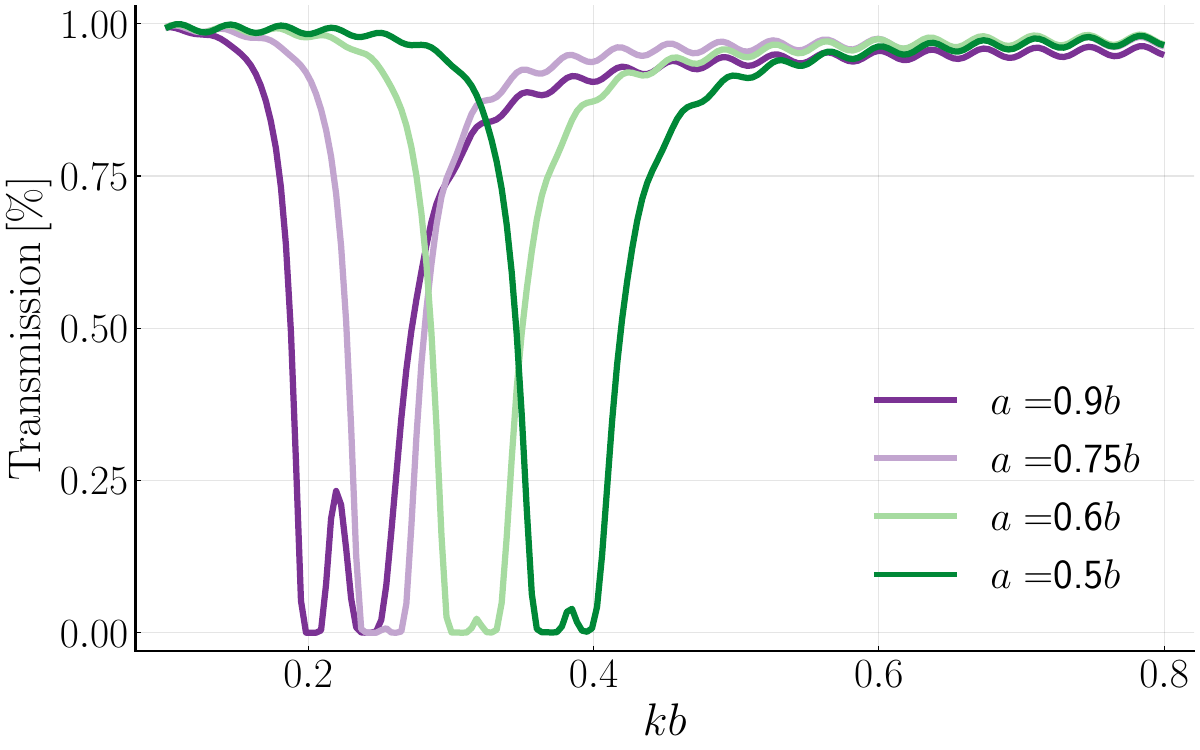}
    \end{subfigure}
    \vspace{-1.1cm}

    (a) \hspace{0.46\textwidth} (b) \hspace{0.4\textwidth}
    \caption{Both graphs show the transmission coefficient through a metamaterial of width $W =$ 32mm filled with randomly positioned resonators, all with the same outer radius $b$. The toal volume fraction of resonators is $\varphi = 12\%$, with a third of them having an aperture size of $\ell_1 = 0.05b$, another third with $\ell_2 = 0.01b$, and the final third with $\ell_3 = 0.005b$. For each curve we use a different inner radius $a$ for all the resonators.}
    \label{fig:three-species-example}
\end{figure}

The transmission results for different values of the inner radii $a$, which changes the wall thickness, are shown in \Cref{fig:three-species-example} below. In \Cref{fig:three-species-example}a we see that as $a$ gets smaller, and the wall thickness increases, the resonant frequency gets lower, however there is a limit to this effect.  In \Cref{fig:three-species-example}b we see that a further decrease in $a$ makes the resonant frequency become higher again, as well as lessening the strength of the resonance. These effects illustrate the need for our formulae to guide the design process.



\subsection{Monte-Carlo validation}
\label{subsec:average_response}

When dealing with disordered materials it is often impractical to predict the precise wave response of the material for each possible configuration of the microstructure. To reach our effective formulae \eqref{eq:effective_formulae} we used ensemble averaging techniques  \cite{Foldy1945}, which lead to practical and explicit results. In this section we validate these formulae by using high-fidelity Monte-Carlo simulations for both a layer and circle filled with resonators. We also highlight some advantages and disadvantages of disordered metamaterials in general.


To get an exact match with a high-fidelity simulation, we need to perform simulations of wave scattering from one configuration at time, where the resonators do not overlap and their positions  are randomly chosen, see \cite{karnezis2024calculating} for details. The average wave is then calculated by taking the average total wave over all simulations. This is what we mean by a high-fidelity Monte-Carlo simulation. The mathematical details on ensemble averaging and Monte-Carlo method are given in \Cref{subsec:ensemble_averaging}; here we focus mostly on showing the results.

\subsubsection*{A layer filled with resonators}
\begin{figure}[H]
    \centering
    \begin{subfigure}[t]{0.45\textwidth}
        \centering
        \adjincludegraphics[width=\linewidth,trim={{.25\width} 0 0 0},clip]{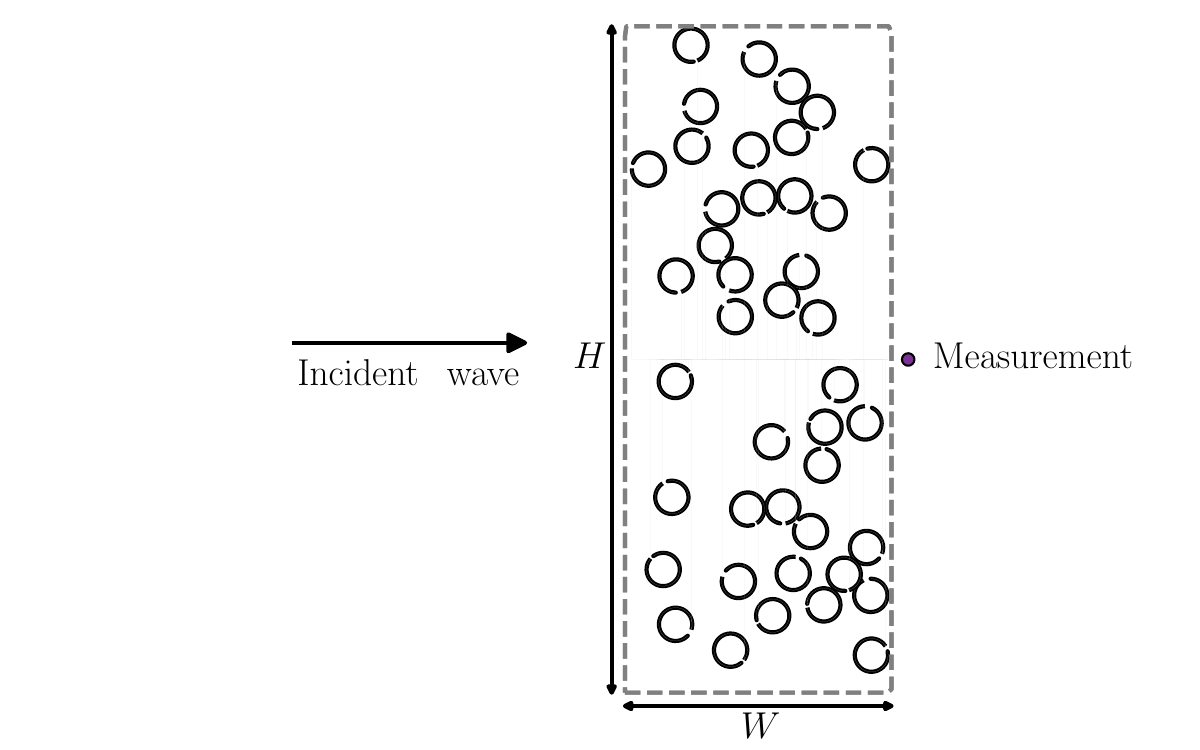}
    \end{subfigure}%
    ~ 
    \begin{subfigure}[t]{0.55\textwidth}
        \centering
        \adjincludegraphics[width=\linewidth]{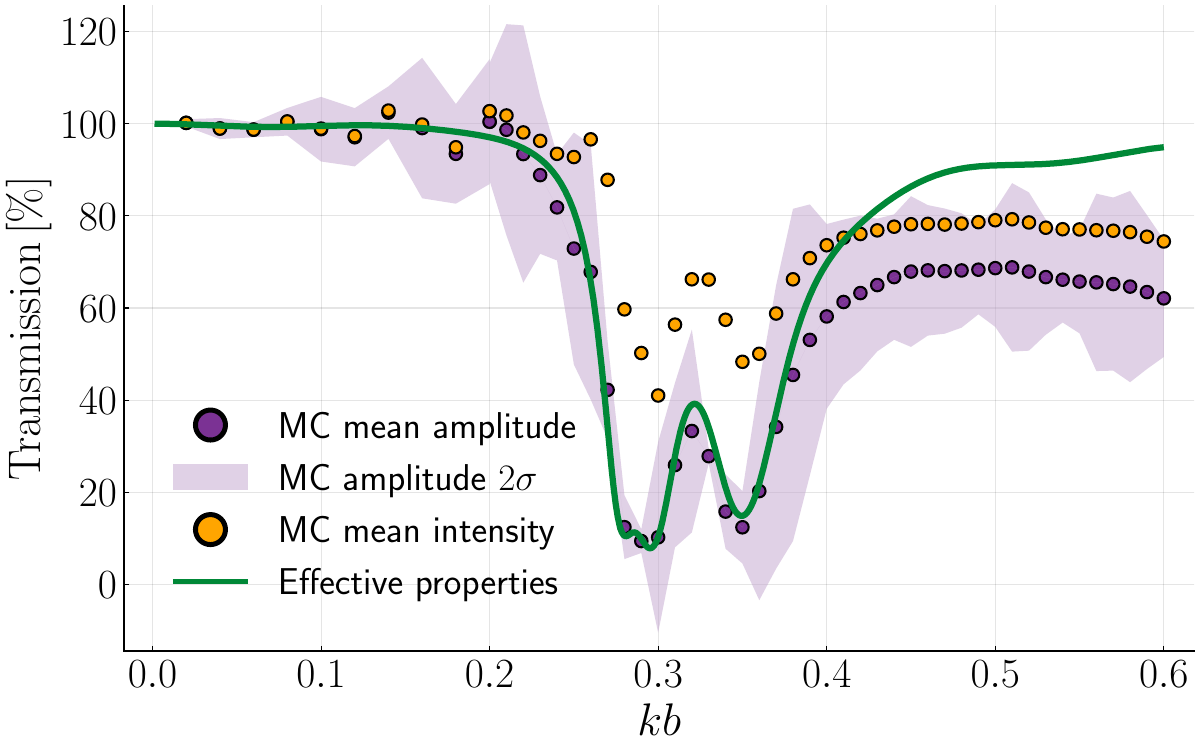}
    \end{subfigure}
    \vspace{-0.85cm}

    (a) \hspace{0.32\textwidth} (b) \hspace{0.38\textwidth}
    \caption{(a) shows one realisation used for our Monte-Carlo (MC) simulations, where resonators are placed within a layer of width $W = 20b$, height $H = 200b$, and $b$ is the resonator radius. An incident plane wave comes from the left and the total field is then measured at a point at half height and one radius away from the layer. This simulation is repeated many times, each with a different configuration of resonators, and then the average MC field is compared with a transmission coefficient calculated by using the effective-properties \eqref{eq:effective_formulae}, shown on the right (b).
    For the MC we use three types of thin-walled resonators, each with 4\% volume fraction, but with different aperture sizes $\ell_1 = 0.05b$, $\ell_2 = 0.01b$, and $\ell_3 = 0.005b$. In total 5,000 different configurations are calculated for the MC results in (b), and $\sigma$ is the 2$\times$ the standard deviation.}
    \label{fig:Monte-Carlo_first_emaple}
\end{figure}

We begin with the case of a layer filled with resonators and an incident plane wave. This is a case which is simpler to understand, though Monte-Carlo simulations can be challenging as the layer needs to be very tall, large $H$ in \Cref{fig:Monte-Carlo_first_emaple}, as the effective theory assumes the layer is infinitely tall.

We perform a high-fidelity Monte-Carlo simulation for a plane incident wave, with wavenumber $k$, scattered by randomly distributed and randomly oriented split-ring resonators ($b = a$ in \Cref{fig:split-ring_resonator}a) inside a long strip shown in \Cref{fig:Monte-Carlo_first_emaple}a. The transmitted field is measured one radius away from the layer as also illustrated in \Cref{fig:Monte-Carlo_first_emaple}a. This simulation is repeated many times, and the average transmission is compared with the prediction from the effective-properties formulae \eqref{eq:effective_formulae} in \Cref{fig:Monte-Carlo_first_emaple}b below. The runtime of the MC simulation was over 24 hours, with parallelization, and using all computational power available, while the effective formulas are computed in a few seconds, using only a fraction of the computational power.

In \Cref{fig:Monte-Carlo_first_emaple}b we see that the transmission calculated using the effective-properties \eqref{eq:effective_formulae} accurately matches the average from Monte-Carlo (MC) in the frequency range $kb \lesssim 0.4$ which includes the resonant frequencies\footnote{Note that some MC points between $kb = 0.08$ and $kb = 0.22$ are higher than 100$\%$ due to diffraction from the corners of the layer, given it has a finite height.}, which agrees with our derivations in \Cref{subsec:low_freq}. The MC mean intensity is higher than the MC mean amplitude, specially around resonance, due to there being no phase cancellation when calculating the mean intensity. Another way to see this is that, in each Monte-Carlo realisation, part of the field has become incoherent or speckled due to scattering \cite{sheng2007introduction}. This incoherent part is lost when taking the average of the field, due to phase cancellation, but not in the average intensity.

%
%


An important feature from \Cref{fig:Monte-Carlo_first_emaple}b is that each realisation of the Monte-Carlo simulation is close to the MC mean amplitude. That is, the purple shaded area has a 95$\%$ statistical confidence (2$\sigma$). This suggests a powerful design strategy: 1) first use the effective formulae to decide on what types of resonators to use, given some target band diagram, then 2) produce one realisation with the resonators randomly placed, and finally 3) perform a local optimisation to adjust the position of the resonators to further refine the transmission or reflecting properties. This design strategy would be much less computationally intensive than pure optimisation strategies used in the literature \cite{Kumar2020,Mao2023,Piggott2015}, as we discussed in \Cref{sec:intro}.

\subsubsection*{A circle filled with resonators}

\begin{figure}[ht!]
    \centering
    \begin{subfigure}[t]{0.45\textwidth}
        \centering
        \adjincludegraphics[width=0.84\linewidth,trim={{.15\width} 0 {.15\width} 0},clip]{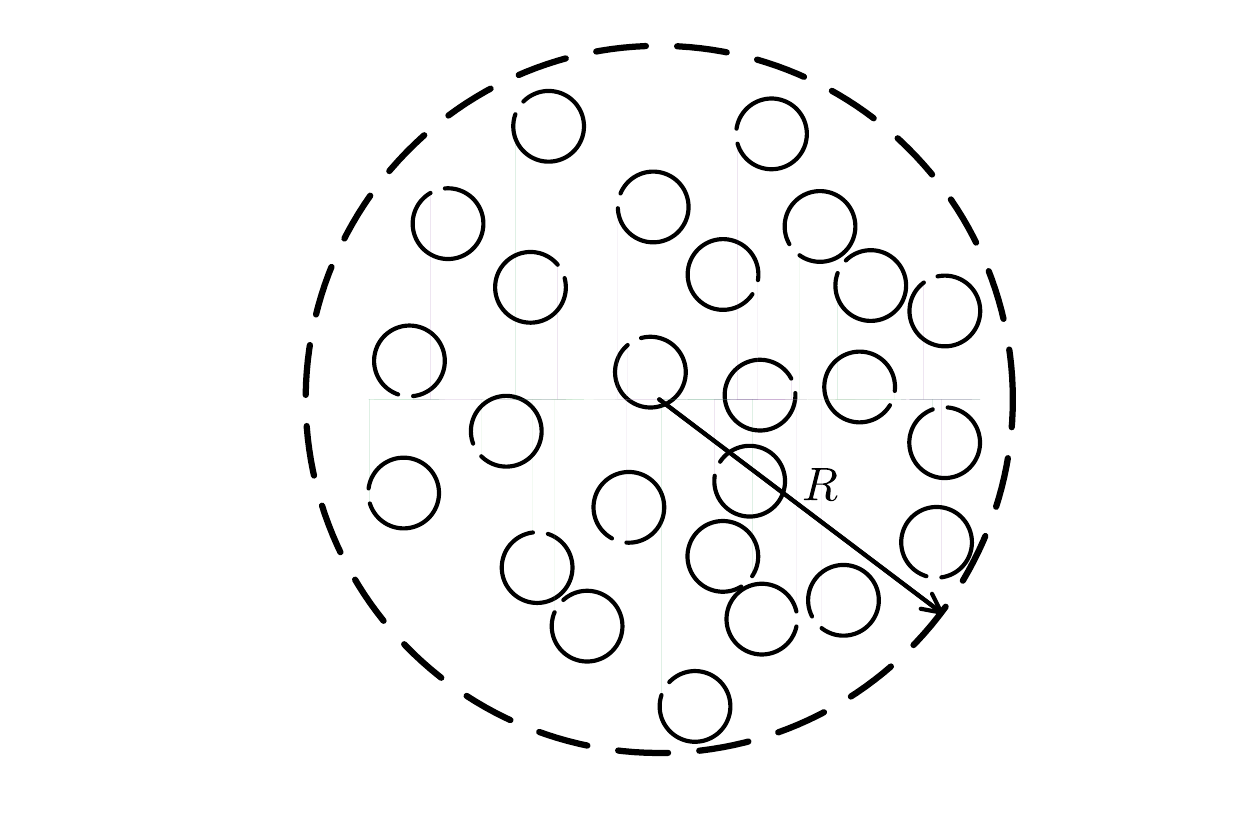}
    \end{subfigure}%
    ~ 
    \begin{subfigure}[t]{0.55\textwidth}
        \centering
        \includegraphics[width=\linewidth]{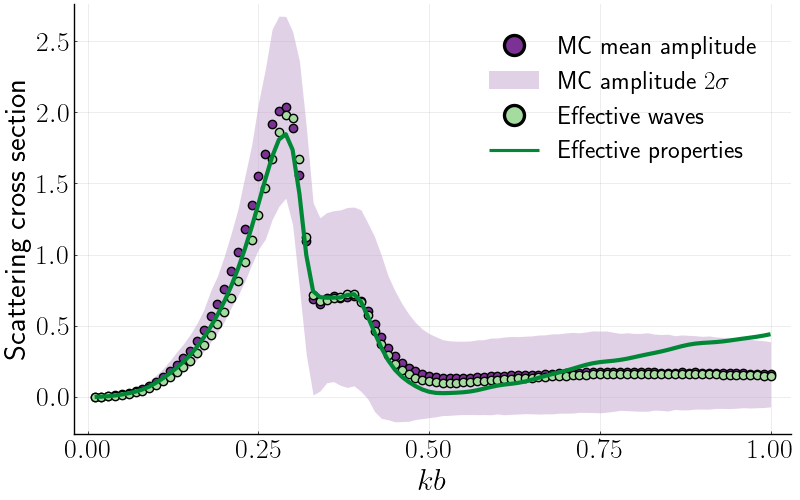}
    \end{subfigure}
        \vspace{-0.86cm}

    (a) \hspace{0.29\textwidth} (b) \hspace{0.38\textwidth}
    \caption{ (a) shows one configuration of resonators used for our Monte-Carlo (MC) simulations. The resonators are placed randomly within a circle of radius $R = 20b$. An incident planar wave comes from the left and the resulting scattering cross section is shown in (b) on the right for three different methods: MC simulations, the effective waves method \cite{Gower_2021}, and by using the effective-properties \eqref{eq:effective_formulae}. The aperture size of all resonators is $\ell = 0.05b$, the volume fraction is $\varphi = 10\%$ in all 40,000 configurations of the MC simulations, and $\sigma$ is the standard deviation${}^2$. }
    \label{fig:Monte-Carlo_second_emaple}
\end{figure}

Our effective formulae \eqref{eq:effective_formulae} are valid for materials of any shape \cite{Gower_2021,gower2023model}. In particular, a circle filled with resonators leads to finite size Monte Carlo simulations (MC) \cite{gower2023model,napal2023effective} which makes validation far simpler. See \Cref{fig:Monte-Carlo_second_emaple}a for an illustration.

Our goal here is to compare three different methods to calculate the scattering cross section of a circle filled with resonators: 1) MC simulations, 2) a circle with the effective properties \eqref{eq:effective_formulae}, and 3) the effective waves methods \cite{Gower_2021,gower2023model} which works beyond low frequencies. The results are shown in \Cref{fig:Monte-Carlo_second_emaple}b. The runtime of the MC simulation was over 5 hours, with parallelization, and using all computational power available. The effective waves method and effective formulas are computed in over 40 minutes and a few seconds respectively, using only a fraction of the computational power.

When performing MC for a circle filled with resonators, we make use of the rotational symmetry to greatly reduce the computational cost of calculating the scattering cross section, see \cite{napal2023effective} for details. For the circle with either effective properties, or using the effective waves methods \cite{Gower_2021,gower2023model}, the scattering cross section becomes:
\begin{equation}
    \label{eq:scat_X_sec}
    \langle \Sigma_{\text{sc}} \rangle = \frac{2}{kR} \sum_{n = -\infty}^{\infty} |\langle \mathcal F_n \rangle|^2,
\end{equation}
where $R$ is the radius of the metamaterial, and $\langle \mathcal F_n \rangle$ are the average material coefficients defined in \Cref{subsec:ensemble_averaging}.

%
%

The results in \Cref{fig:Monte-Carlo_second_emaple}b again show that the mean of MC data is close to the results using the simple formulae for the effective-properties at lower frequencies $kb \lesssim 0.7$. The effective-properties are deduced as a low-frequency asymptotic approximation of the effective waves method, and it can be seen in \Cref{fig:Monte-Carlo_second_emaple} that these two methods match for lower frequencies. We can also see that the effective waves method matches the Monte-Carlo results for all frequencies. However, the effective waves method has a significant drawback: it requires a much more elaborate calculation \cite{gower2023model} than the formulae \eqref{eq:effective_formulae}, which makes it less useful than the effective formulae for designing metamaterials. 


\footnotetext[2]{The scattering cross section \eqref{eq:scat_X_sec} was calculated with the first nine scattering coefficients $\langle \mathcal F_n \rangle$ in \eqref{eq:scat_X_sec}.}

\section{Multiple scattering and ensemble average}
\label{sec:methods}

Here we derive the effective-properties \eqref{eq:effective_formulae} in three steps: scattering by a single resonator in \Cref{subsec:single_scattering}, scattering of the whole metamaterial in \Cref{subsec:ensemble_averaging}, and then an asymptotic low-frequency expansion in \Cref{subsec:low_freq}.

\subsection{Scattering by a single resonator}
\label{subsec:single_scattering}

Before calculating the acoustic response of the whole metamaterial, we need to describe how each resonator scatters waves.
This is best done with the T-matrix method \cite{Waterman_1,Waterman_2,martin2006multiple}, which we describe below, as it facilitates calculations for multiple scattering. 

Let the centre of the resonator be the origin of $\mathbb R^2$. We assume the incident wave is a regular function, which allows us to expand the incident wave in terms of a series of regular radial waves:
\begin{equation}
    \label{eq:regular_expansion}
    u_{\text{inc}}(\bm{r}) = \sum_{n = -\infty}^{\infty} g_n \mathrm V_n(k \bm{r}) \quad \text{with} \quad
      \mathrm V_n(k \bm{r}) = \mathrm J_n(k r) \, \mathrm e^{\mathrm i n \theta},
\end{equation}
where $\bm{r} \in \mathbb R^2$, $k$ is the wavenumber, 
%
$\mathrm J_n$ being the Bessel function of the first kind, and $(r,\theta)$ polar coordinates of $\mathbb R^2$. Note that the time-harmonic dependence $\mathrm{e}^{-\mathrm i \omega t}$ is assumed throughout the paper for both incident and scattered waves, where $\omega$ is the angular frequency, and $k = \omega / c$, where $c$ is the speed of sound in the background medium.

Due to the linearity of the Helmholtz equation, the total field is given by the superposition of the incident wave and the scattered field from our resonator:
\begin{equation}
    \notag
    u_{\text{tot}}(\bm{r}) = u_{\text{inc}}(\bm{r}) + u_{\text{sc}}(\bm{r}),
\end{equation}
and, outside the resonator, the scattered field can be represented as a series of outgoing radial modes:
\begin{equation}
    \label{eq:scattered_field}
    u_{\text{sc}}(\bm{r}) = \sum_{n = -\infty}^{\infty} f_n \mathrm U_n (k \bm{r}), \quad \text{for} \quad r = |\bm{r}| > b,
\end{equation}
where $b$ is the outer radius of the resonator, $f_n$ are the scattering coefficients, and the outgoing radial modes are given by:
\begin{equation}
\notag
    \mathrm U_n(k \bm{r}) = \mathrm H_n(k r) \, \mathrm e^{\mathrm i n \theta},
\end{equation}
with $\mathrm H_n$ being the Hankel function of the first kind\footnote{The usual superscript $(1)$ on the Hankel function is suppressed here and henceforth for clarity.}.

To determine the scattered field, we need to apply boundary conditions at the walls of the resonator. The result of solving the boundary conditions can be expressed in terms of the T-matrix which relates the incident wave to the scattered waves through:
%
\begin{equation}
    \label{eq:T-matrix}
    f_n = \sum_{m = -\infty}^{\infty} T_{n m} g_m,
\end{equation}
where $T_{n m}$ are the elements of the T-matrix. For a circular Helmholtz resonator with sound-hard walls (Neumann boundary conditions) and outer radius $b$, we deduce in \Cref{app:T-matrix} the T-matrix:
\begin{equation}
    \label{eq:T-matrix_general}
    \tcboxmath{
    T_{np} = - \frac{\mathrm J'_p(kb)}{\mathrm H'_p(kb)}\delta_{np} - \frac{\mathrm i \mathrm e^{-\mathrm i(n - p) \theta_{0}}}{\pi \mathrm H'_n(kb)\mathrm H'_p(kb)} z(\lambda)
    },
\end{equation}
where $\theta_0$ is the orientation of the aperture (anticlockwise angle with the $x$-axis), prime denotes differentiation with respect to the argument, and $\lambda$ is a set of properties that identifies one type of resonator. 

\begin{figure}[ht]
    \centering
    \adjincludegraphics[width=0.6\linewidth,trim={{.05\width} {.35\height} 0 {.3\height}},clip]{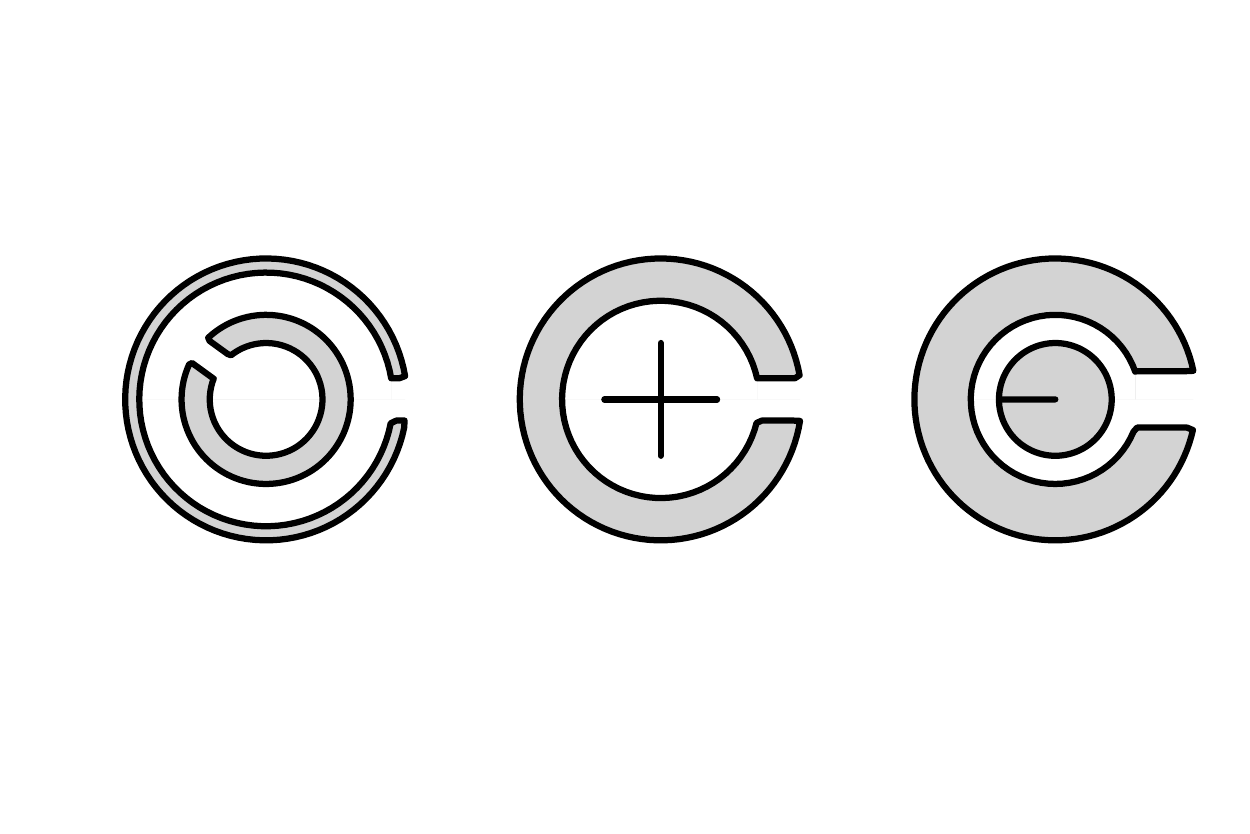}
    \caption{Three examples of Helmholtz resonators with different resonance factors $z(\lambda)$.}
    \label{fig:possible_resonators}
\end{figure}

In \Cref{app:T-matrix} we use a closed form for $z(\lambda)$ for the split-ring resonator in \Cref{fig:split-ring_resonator} which is fully characterised by $\lambda = \{ kb, ka, k\ell\}$, with outer radius $b$, inner radius $a$, and aperture of $\ell$. However, the formula \eqref{eq:T-matrix_general} is the same for any circular resonant structure, with only $z(\lambda)$ changing, with some examples illustrated in \Cref{fig:possible_resonators}. 


To determine the resonance factor $z(\lambda)$ for a different resonator, one should redo the asymptotic calculations in \cite{Atom-Helmholtz} for its specific internal geometry, and then repeat the deduction in \Cref{app:T-matrix} to arrive at \eqref{eq:T-matrix_general}. Similarly, we could easily adapt \eqref{eq:T-matrix_general} to incorporate the case of multiple apertures in the same resonator.

\subsection{Ensemble average and multiple scattering}
\label{subsec:ensemble_averaging}

To account for different types of resonators in the metamaterial we need to first calculate the exact multiple scattering between the resonators, after which we will perform ensemble averaging as shown in \cite{Gower_2021,napal2023effective,martin2006multiple}. We also borrow the notation from these references.

Let us number each resonator in the material. We denote the centre of the $j$-th resonator by $\bm{r}_j$, and the set of its properties by $\lambda_j$. As an example, if the $j$-th resonator is a split-ring resonator, shown in \Cref{fig:split-ring_resonator}, we have the following set of properties:
\begin{equation}
    \notag
    \lambda_j = \{k b_j,k a_j,k \ell_j\}
\end{equation}
where $b_j$ is the outer radius, $a_j$ the inner radius, and $\ell_j$ the aperture size.

The scattered field from all resonators is a sum of the waves scattered from each resonator:
\begin{equation}
    \label{eq:complete:scattered_wave}
    u_{\text{sc}} (\bm{r}) = \sum_{j=1}^J \sum_{n = -\infty}^{\infty} f_n^j \mathrm U_n(k\bm{r} - k\bm{r}_j), 
\end{equation}
where $\bm r$ can not be within any resonator, $J$ is the total number of resonators, and $f_n^j$ are the scattering coefficient of the $j$-th resonator as introduced in \Cref{subsec:single_scattering}. 

For numerical validation, and to help explain the ensemble averaging, it helps to specialise to the case of all resonators within a large sphere. To achieve this we use Graf's addition theorem, to rewrite \eqref{eq:complete:scattered_wave} in terms of outgoing waves centred as the origin:
\begin{equation}
    \label{eq:material_scattered_wave}
    u_{\text{sc}} (\bm{r}) = \sum_{j=1}^{J} \sum_{n,n' = -\infty}^{\infty} \mathrm V_{n - n'}( - k\bm{r}_j) f_n^j (\Lambda) \mathrm U_{n'}(k\bm{r}).
\end{equation}
The coefficients $f_n^j$ depend on all properties and positions of all the resonators \cite{Gower_2021, martin2006multiple}. To make this explicit, and simplify notation, we denote one configuration by $\Lambda$, which represents all the positions and properties of the resonators. We also define the material scattering coefficients of the whole sphere containing all the resonators as:
\begin{equation}
    \notag
    \mathcal F_{n} (\Lambda) = \sum_{j=1}^{J} \sum_{n' = -\infty}^{\infty} \mathrm V_{n' - n}( - k\bm{r}_j) \, f_{n'}^j (\Lambda),
\end{equation}
which simplifies \eqref{eq:material_scattered_wave} into:
\begin{equation}
    \label{eq:material_scattering_coefficients}
    u_{\text{sc}} (\bm{r}) = \sum_{n' = -\infty}^{\infty} \mathcal{F}_{n} (\Lambda) \mathrm U_{n}(k\bm{r}).
\end{equation}

To calculate the average response of the metamaterial, we ensemble average over all possible positions, orientations and properties \cite{Gower_2021,gower_smith_parnell_abrahams_2018, Foldy1945}, while assuming that resonators do not overlap, and that any possible configuration $\Lambda_i$ has the same probability (micro-canonical ensemble), which results in:
\begin{equation}
    \label{eq:average_material_response}
    \langle u_{\text{sc}} (\bm{r}) \rangle = \frac{1}{M} \sum_{i=1}^M \sum_{n = -\infty}^{\infty} \mathcal F_n(\Lambda_i) \mathrm U_{n}(k\bm{r}) = \sum_{n = -\infty}^{\infty} \langle \mathcal{F}_{n} \rangle \mathrm U_{n}(k\bm{r}),
\end{equation}
where the bracket notation $\langle \, \circ \, \rangle$ denotes the ensemble average of $\circ$, and $M$ is the number of possible configurations $\Lambda_i$ which tends to infinite.

The average material coefficients $\langle \mathcal F_n \rangle$ can be calculated either by the effective waves method \cite{Gower_2021,napal2023effective}, or by brute force Monte-Carlo simulations using the Julia library MultipleScattering.jl \cite{2020MultipleScatering.jl}.

The effective waves method calculates the average amplitude \eqref{eq:average_material_response} directly, avoiding the explicit computation of the scattering from each configuration \eqref{eq:material_scattering_coefficients}. Further, this method provides a semi-analytic formula for the dispersion equation of the meta-material, which we use to obtain a low-frequency expansion in \Cref{subsec:low_freq}, resulting in explicit formulae for the effective-properties.

On the other hand, Monte-Carlo simulations estimate the average response by explicitly calculating the scattering from each configuration $\Lambda_i$, and then repeating this for a very large number of possible configurations, and then calculate the average \eqref{eq:average_material_response}. This strategy also allows the computation of higher statistical moments of the scattered field, such as the mean scattered intensity:
\begin{equation}
    \label{eq:transmitted-intensity}
    \langle | u_{\text{sc}}(\bm{r}) |^2 \rangle = \sum_{n,m = -\infty}^{\infty} \langle \mathcal{F}_{n}^* \mathcal{F}_{m} \rangle \mathrm U_{n}^*(k\bm{r}) \mathrm U_{m}(k\bm{r}),
\end{equation}
where $*$ means complex conjugation. In \Cref{subsec:average_response} we used both the mean amplitude \eqref{eq:average_material_response} and mean intensity \eqref{eq:transmitted-intensity} calculated from Monte-Carlo simulations for our disordered metamaterial, and compare them against the average response from the effective-properties \eqref{eq:effective_formulae} deduced in \Cref{subsec:low_freq} that follows.

\subsection{low-frequency expansion}
\label{subsec:low_freq}

In this section, we use the dispersion equation for a 2D random particulate material from the effective waves method in \cite[equation 4.10]{napal2023effective} to derive effective-properties of the metamaterial with resonators. As we are interested in the long wavelength regime, we assume that the resonators' positions, orientations, and properties are uncorrelated, except that the resonators can not overlap\footnote{To reach a dispersion equation we also need a closure assumption. We use the Quasi-Crystalline Approximation discussed in \cite{piva2024,gower_smith_parnell_abrahams_2018}.}. For these assumptions, the dispersion equation \cite[equation 4.10]{napal2023effective} simplifies into:
\begin{equation}
    \label{eq:dispersion_equation}
    F_{m}(\lambda_1) + \sum_{n' = -\infty}^{\infty} \frac{2 \pi \overline{T}_{n}(\lambda_1)}{k_\star^2 - k^2} \int \mathrm N_{n'-m} [k (b_{1} + b_2), k_\star (b_{1} + b_{2})] F_{n'}(\lambda_2) \mathfrak{n}(\lambda_2) \mathrm d \lambda_2 = 0,
\end{equation}
where $\mathrm N_{n}[x,y] = x \mathrm H'_n(x) \mathrm J_n(y) - y \mathrm J'_n(y)\mathrm H_n(x)$, $\mathfrak n(\lambda)$ is the number density of resonators of type $\lambda$, $k_\star$ is the effective wavenumber, $\overline{T}_{n}(\lambda)$ is the isotropic T-matrix \eqref{eq:diagonal_T-matrix} derived in \Cref{app:T-matrix}, and $F_n$ is the amplitude of the effective wave, defined in \cite{napal2023effective}. 

We define the effective speed of sound as follows:
\begin{equation}
    \notag
    c_* = c k / k_\star.
\end{equation}
To do an asymptotic expansion, we assume there is a maximum outer radius for the resonators $b_\text{max}$, such that $b_j \leq b_{\text{max}}$ for all $j$. We also recall that all resonators are sub-wavelength, so the dimensionless wavenumber $\epsilon = k b_{\text{max}}$ is small. Then, we expand both the speed of sound and effective wavenumber as a power series of $\epsilon$:
\begin{equation}
    \label{eq:effective_speed_expansion}
    \frac{c}{c_*} = \chi + \mathcal O (\epsilon) \quad \text{and} \quad k_\star b_j = \chi \alpha_j \epsilon + \mathcal O(\epsilon^2),
\end{equation}
where $\alpha_j = b_j / b_{\text{max}}$, and $\chi$ has yet to be determined. 

To determine $\chi$ we need to do an asymptotic expansion of the terms in \eqref{eq:dispersion_equation}, including $\overline{T}_{n}(\lambda_1)$ which depends on the resonance factor. The trick to do this elegantly is to make no assumptions about the asymptotic order of $z(\lambda)$. This is because  $z(\lambda)$ depends on the frequency in a non-trivial way, involving all parameters of the resonators internal geometry (see \eqref{eq:int_geom_factor} and \eqref{eq:h_expression} in \Cref{app:T-matrix,app:expansion}). Fortunately we can reach simple formulae without making any assumption about $z(\lambda)$. Instead, the results error of our method will be relative to $z(\lambda)$.

Performing an asymptotic expansion on the terms in \eqref{eq:dispersion_equation} results in:
\begin{equation}
    \label{eq:leading_order_eigensystem_terms}
    \begin{aligned}
        &\frac{1}{k_\star^2 - k^2} = \frac{b_{\text{max}}^2}{\chi^2-1} \frac{1}{\epsilon^2} + \mathcal O(\epsilon^{-1}),
\\
        &\mathrm N_{n'-m}(k b_{1,2}, k_\star b_{1,2}) = \mathrm N_{n'-m}^{(0)}(\chi) + \mathcal O(\epsilon), 
        \quad \text{with} \;\; \mathrm N_{n}^{(0)}(\chi) = \frac{2 \mathrm i}{\pi} \sum_{m = -\infty}^{\infty} \chi^{|m|} \delta_{n,m},
\\
        &\overline{T}_{n}(\lambda_1) = \overline{T}_{n}^{(2)}(\lambda_1) \alpha_1^2 \epsilon^2 + \mathcal O(\epsilon^3( 1 + z(\lambda))), \quad \text{with} 
        \\
        &\overline{T}_n^{(2)} (\lambda_1) = \frac{\mathrm i \pi}{4} \left[ \delta_{n,1} + \delta_{n,-1} - \delta_{n,0} + z(\lambda_1) \delta_{n,0} \right].
    \end{aligned}
\end{equation}
where $\delta_{n, m}$ is the delta Kronecker symbol. We have included $z(\lambda)$ in the error so that we do not need to make assumptions about the size of $z(\lambda)$ which can be large. Fortunately, this works well, as the maximum relative error of the leading order term for $\overline{T}_{n}$ is $\epsilon$. We see this trend in the numerical validation.


 Substituting \eqref{eq:leading_order_eigensystem_terms} into \eqref{eq:dispersion_equation}, and retaining only the leading order terms we reach: 
\begin{equation}
    \label{eq:leading_order_eigensystem}
    F_{n}(\lambda_1) - \frac{2 \pi b_{\text{max}}^2}{1 - \chi^2} \overline{T}_n^{(2)}(\lambda_1) \alpha_1^2 \sum_{n' = -1}^1 \mathrm N_{n'-n}^{(0)}(\chi) \int F_{n'}(\lambda_2) \mathfrak{n}(\lambda_2) \mathrm d \lambda_2 = 0,
\end{equation}
where the sum is truncated to three terms as the other terms are of higher order in $\epsilon$. To determine $\chi$ we multiply both sides of \eqref{eq:leading_order_eigensystem} by $\mathfrak{n}(\lambda_1)$ and integrate over $\lambda_1$, leading to the following eigenvalue problem:
\begin{equation}
    \label{eq:eigen_system}
    \begin{aligned}
        &\sum_{n' = -1}^1 \mathrm M_{n n'} \tilde{F}_{n'} = 0,
        \quad \text{with} \quad \tilde{F}_{n} = \int F_{n}(\lambda_1) \mathfrak n(\lambda_1) \mathrm d \lambda_1 \quad \text{and}
\\
        &\mathrm M_{n n'} = \delta_{n,n'} - \frac{2 \pi b_{\text{max}}^2}{1 - \chi^2} \mathrm N_{n'-n}^{(0)}(\chi)  \int \overline{T}_n^{(2)} (\lambda_1) \alpha_1^2 \mathfrak n(\lambda_1) \mathrm d \lambda_1.
    \end{aligned}
\end{equation}
By solving the dispersion relation $\det \mathrm M = 0$, we can obtain an exact expression for $\chi$ which substituted into \eqref{eq:effective_speed_expansion} leads to (at leading order):
\begin{equation}
    \label{eq:effective_speed}
    \tcboxmath{
    c_*^2 = \frac{\beta}{\rho} \frac{ (1 - \varphi)}{ \left(1 - \varphi + 
\sum_{j=1}^N z(\lambda_j) \phi_j \right) (1 + \varphi)}
    },
\end{equation}
where, to simplify the exposition, we have specialised the above to a mixture of $N$ types of resonators which led us to substitute 
\begin{equation}
    \label{eq:number_density}
    \mathfrak{n}(\lambda) = \sum_{j=1}^N \frac{\phi_j}{\pi b_j^2} \delta(\lambda - \lambda_j),
\end{equation}
where $\phi_j$ is the volume fraction of the resonator of type $j$, $\varphi = \sum_{j=1}^N \phi_j$ is the total volume fraction of resonators, and $\delta$ is the Dirac delta distribution. 


Finally, we want an effective density $\rho_\star$ and bulk modulus $\beta_\star$ such that $c_\star^2 = \beta_\star / \rho_\star$. Analogous to \cite{Effective_density,Gower_2021} there is only one way to factor out $\beta_\star$ and $\rho_\star$ such that the limits $\phi_j \to 0$ are consistent, and they lead to \eqref{eq:effective_formulae} presented in \Cref{sec:design}. Then, to calculate the average response of any metamaterial, one can replace it with an effective homogeneous medium with properties given by \eqref{eq:effective_formulae}. We validate these results against Monte-Carlo simulations in \Cref{fig:Monte-Carlo_first_emaple,fig:Monte-Carlo_second_emaple}.

We note that it is expected that Helmholtz resonators, with only one gap such as the examples shown in \Cref{fig:possible_resonators}, only significantly alter the effective bulk modulus, as we deduced in \eqref{eq:effective_formulae}. This is because one gap leads to a dominant monopole term, and the monopole terms from the resonators contribute to the effective bulk modulus while the dipole terms contribute to the effective density. Therefore a dipole dominant resonator would lead to a significant change in the effective density.


\section{Concluding remarks and further avenues}
\label{sec:conclusion}

\textbf{Summary.} We have introduced a disordered composite metamaterial, consisting of sub-wavelength Helmholtz resonators. Having small resonators compared to the wavelength ($kb<1$) allowed us to deduce, from first principles, formulae for both effective bulk modulus and effective mass density \eqref{eq:effective_formulae}. These formulae were validated against high-fidelity Monte-Carlo simulations for both a layer and a circle filled with resonators. Using our effective formulae, we are able to quickly design broad frequency band gaps without using multiple layers, periodicity, or heavy optimisation methods. In future work, we plan to expand our results to higher order in $kb$ and hence produce formulae that hold at higher frequencies, allow for effects of differing distributions of inclusions, and allow us quantify the difference between, say, random, periodic and hyperuniform materials.



\noindent \textbf{Ensemble average properties.} Contrary to the standard approach with periodic media, the formulae we deduce are based on ensemble averaging over the possible positions and orientations of the resonators, which introduces both advantages and disadvantages. The key advantages are that: 1) we were able to easily deduce effective-properties for any mix of different types of resonators, and 2) the effective properties are robust with respect to changes in position and orientation of the resonators, as shown by the motivation example in \Cref{fig:all_mighty_frequency_splitter}. However, the effective properties in \eqref{eq:effective_formulae} do not give exactly the same results as any one specific configuration, and they also do not capture the average intensity of the scattered waves.

\noindent \textbf{Applications.} Despite using ensemble averaging in deriving \eqref{eq:effective_speed}, there are many applications for a single configuration of resonators. For example, the frequency demultiplexer in \Cref{fig:all_mighty_frequency_splitter} with only one configuration of resonator generated randomly had a similar performance compared with the one achieved via heavy optimisation in \cite{Piggott2015}. In part this is due to the response of a single configuration being close to the mean response for low frequencies, as illustrated by the purple shaded region in \Cref{fig:Monte-Carlo_first_emaple,fig:Monte-Carlo_second_emaple} which represents two standard deviations of the mean.

\noindent \textbf{More general resonators.} Throughout the whole paper we only showed examples of split-ring resonators from \cite{Atom-Helmholtz,Dave's-latticeI,Dave's-latticeII}. However, the formulae for the T-matrix \eqref{eq:T-matrix_general} has the same form for any sub-wavelength circular Helmholtz resonator (which has one aperture), like the ones depicted in \Cref{fig:possible_resonators}, and the derived effective-properties \eqref{eq:effective_formulae} would have the same formula, although the resonance factor $z(\lambda)$ would change. Resonators with different internal structures should result in richer effective-properties for the metamaterial, as illustrated in \Cref{fig:three-species-example} for thick walled split-ring resonators.

\noindent \textbf{Optimising band-gaps.} One interesting direction to explore as future work is to implement the enhanced design strategy described in \Cref{subsec:average_response}. This strategy consists of finding one specific configuration for the particles via optimisation while being guided by the overall band structure from the effective-properties \eqref{eq:effective_formulae}.

\noindent \textbf{Future generalisations.} Another future avenue would be to investigate possible generalisations of the metamaterial presented. The simplest example is the three-dimensional case, where the Helmholtz resonators are spherical shells with a small aperture. Different from the case of long cylinders, small spheres could be used to produce compact versions of the metamaterial presented in three dimensions. Other than just acoustics, the case of sub-wavelength resonating structures in electromagnetism or elasticity could lead to interesting applications. For example, one could study how to embed sub-wavelength resonators in the building blocks of low-frequency operating machinery or other structures, to prevent harmful vibrations from propagating.

\section*{Acknowledgement}

The author(s) would like to thank the Isaac Newton Institute for Mathematical Sciences, Cambridge, for support and hospitality during the programme Mathematical Theory and Applications of Multiple Wave Scattering, where work on this paper was undertaken. This work was supported by EPSRC grant EP/R014604/1.
Paulo Piva gratefully acknowledges funding from an EPSRC Case studentship with Johnson Matthey. Art Gower gratefully acknowledges support from EPSRC (EP/V012436/1). David Abrahams gratefully acknowledges funding from the Royal Society for an Industry Fellowship with Thales UK.

\section*{Appendices}

\appendix

\section{Expression for the T-matrix}
\label{app:T-matrix}

In this section, we deduce the T-matrix for a split-ring Helmholtz in \eqref{eq:T-matrix_general} and use an expression for $z(\lambda)$ in \eqref{eq:int_geom_factor} combined with \eqref{eq:h_expression} for a split ring resonator.

Consider a resonator with outer radius $b$, inner radius $a$, aperture size $2\ell$, and orientation $\theta_0$, as shown in Figure \ref{fig:split-ring_resonator}. Let us obtain the scattered field for the incident plane incident wave:
\begin{equation}
    \notag
    v_{\text{inc}}(\bm{r}) = \mathrm e^{\mathrm i k r \cos (\theta - \theta_{\text{inc}})},
\end{equation}
Using \cite[Equation (3.1)]{Atom-Helmholtz} we can calculate the scattered field for $\theta_0 = 0$ which in our notation becomes
\begin{equation}
    \label{eq:field_outside}
    v_{\text{sc}} (\bm{r}) = A \mathrm U_0(k \bm{\tilde{r}}) +  \sum_{n = -\infty}^{\infty} c_n \mathrm U_n(k \bm{r}),
\end{equation}
where $\bm{\tilde{r}} = \bm{r} - b(\cos(\theta_{0}), \sin(\theta_{0}))$ is the aperture location, $c_n$ and $A$ are coefficients given by \cite[Eqs. (3.4) and (3.6) respectively]{Atom-Helmholtz}. To obtain the scattered field for $\theta_0 \not = 0$, we rotate the coordinate system $\theta \to \theta - \theta_{0}$, followed by the rotation of the incident wave $\theta_{\text{inc}} \to \theta_{\text{inc}} - \theta_{0}$. These rotations leave the incident wave unaltered while rotating the resonator by an angle of $\theta_{0}$ which together with \cite[Equation (3.1)]{Atom-Helmholtz} results in:
\begin{equation}
    \label{eq:plane_wave_coefs}
    \begin{aligned}
        &A = - z(\lambda) \sum_{p = -\infty}^{\infty} \frac{\mathrm i^p kb}{\mathrm H'_p(kb)} \mathrm e^{\mathrm i p (\theta_{0} - \theta_{\text{inc}})},
\\
        &c_n = - \mathrm i^n \frac{\mathrm J_n'(k b)}{\mathrm H'_n (k b)} \mathrm e^{-\mathrm i n \theta_{\text{inc}}} - \frac{A}{2} \frac{\mathrm Q_n(kb)}{\mathrm H'_n(kb)} \mathrm e^{-\mathrm i n \theta_{0}},
    \end{aligned}
\end{equation}
where $\mathrm Q_n (x) = \mathrm J_n(x) \mathrm H'_n(x) + \mathrm J'_n(x) \mathrm H_n(x)$, the prime notation denotes the derivative with respect to the argument. We call the function $z(\lambda)$ the resonance factor and it is defined by \eqref{eq:int_geom_factor} in \Cref{app:expansion}.

To rewrite \eqref{eq:field_outside} in terms of a T-matrix, shown in \eqref{eq:T-matrix}, we need to express all the terms centred at the origin, which leads us to use Graf's addition theorem to rewrite the monopole term evaluated at $k \bm{\tilde{r}}$ in the form:
\begin{equation}
    \notag
    \mathrm U_0(k \bm{\tilde{r}}) = \sum_{n = -\infty}^{\infty} \mathrm J_{-n}(kb) \mathrm e^{\mathrm i n (\pi - \theta_{0})} \mathrm U_n(k\bm{r}),
\end{equation}
which substituted into \eqref{eq:field_outside} leads to 
%
\begin{equation}
    \label{eq:field_outside_one_series}
    v_{\text{sc}} (\bm{r}) = \sum_{n=-\infty}^{\infty} d_n \mathrm U_n(k\bm{r}), \quad \text{with} \quad d_n = c_n + A \mathrm J_n(kb) \mathrm e^{-\mathrm i n \theta_{0}}.
\end{equation}

Next, the T-matrix \eqref{eq:T-matrix} relates any incident wave to the scattered wave, so we need to rewrite any incident wave $u_{\text{inc}}(\bm{r})$ in terms of plane waves to use the results above. To achieve this, we use the Jacobi-Anger expansion of the plane wave:
%
\begin{equation}
    \notag
    v_{\text{inc}}(\bm{r}) = \mathrm e^{\mathrm i k r \cos(\theta - \theta_{\text{inc}})} = \sum_{n = -\infty}^{\infty} \mathrm i^n \mathrm J_n(kr) \mathrm e^{\mathrm i n (\theta - \theta_{\text{inc}})},
\end{equation}
followed by a superposition of plane waves:
\begin{equation}
\label{eq:incident_wave_superposition}
    \begin{aligned}
        u_{\text{inc}}(\bm{r}) &= \int_0^{2\pi} v_{\text{inc}}(\bm{r}) \mathrm g(\theta_{\text{inc}}) \mathrm d \theta_{\text{inc}}
    = \int_0^{2\pi} \sum_{n} \mathrm g(\theta_{\text{inc}}) \mathrm i^n \mathrm J_n(kr) \mathrm e^{\mathrm i n (\theta - \theta_{\text{inc}})} \mathrm d \theta_{\text{inc}}
\\
        &= \sum_{n} \left[ \mathrm i^n \int_0^{2\pi} \mathrm g(\theta_{\text{inc}}) \mathrm e^{-\mathrm i n \theta_{\text{inc}}} \mathrm d \theta_{\text{inc}} \right] \mathrm V_n(k \bm{r}).
    \end{aligned}
\end{equation}
Without loss of generality, we choose the amplitude of the packet of plane waves $\mathrm g(\theta_{\text{inc}})$ in \eqref{eq:incident_wave_superposition} such that:
\begin{equation}
    \notag
    \mathrm i^n \int_0^{2\pi} \mathrm g(\theta_{\text{inc}}) \mathrm e^{-\mathrm i n \theta_{\text{inc}}} \mathrm d \theta_{\text{inc}} = g_n,
\end{equation}
which implies that \eqref{eq:incident_wave_superposition} now matches the form of any regular incident wave \eqref{eq:regular_expansion}.


Now we need to perform the same operations on the scattered field $v_{\text{sc}}$ \eqref{eq:field_outside_one_series} from a plane wave to obtain the total scattered field from any incident wave $u_{\text{sc}}$, which is possible due to the linearity of the Helmholtz equation and results in:
%
\begin{equation}
    \label{eq:superposition_scat_wave}
    \begin{aligned}
        u_{\text{sc}} (\bm{r}) &= \int_{0}^{2\pi} v_{\text{sc}}(\bm{r}) \mathrm g(\theta_{\text{inc}}) \mathrm d \theta_{\text{inc}}
        = \sum_{n} \left[ \int_0^{2\pi} d_n \mathrm g(\theta_{\text{inc}}) \mathrm d \theta_{\text{inc}} \right] \mathrm U_n(k\bm{r}),
\\
        &= \sum_{n,p} \left[ -\delta_{np} \frac{\mathrm J'_p(kb)}{\mathrm H'_p(kb)} + \frac{2 \mathrm e^{-\mathrm i(n - p) \theta_{0}}}{(\pi k b)^2 h(\lambda)\mathrm H'_n(kb)\mathrm H'_p(kb)} \right] g_p \mathrm U_n(k\bm{r}),
    \end{aligned}
\end{equation}
where we have used the Wronskian of Bessel-Hankel functions in the second line:
\begin{equation}
    \notag
    \mathrm J_n(x) \mathrm H'_n(x) - \mathrm J'_n(x) \mathrm H_n(x) = \frac{2 \mathrm i}{\pi x}.
\end{equation}
Finally by comparing \eqref{eq:T-matrix}, \eqref{eq:T-matrix_general} and \eqref{eq:superposition_scat_wave} we conclude that:
\begin{equation}
    \label{eq:T-matrix_expression}
    T_{np} = - \frac{\mathrm J'_p(kb)}{\mathrm H'_p(kb)}\delta_{np} + \frac{\mathrm i \mathrm e^{\mathrm i(n - p) \theta_{0}}}{\pi \mathrm H'_n(kb)\mathrm H'_p(kb) } z(\lambda),
\end{equation}
which is the general T-matrix for a Helmholtz resonator using formulae in \cite{Atom-Helmholtz}.



In \Cref{subsec:low_freq}, the T-matrix \eqref{eq:T-matrix_general} is used to calculate the dispersion equation. However, as discussed in \cite{Gower_2021,napal2023effective}, only the average T-matrix over all orientations of the aperture $\theta_0$ contributes to the dispersion equation. We call this orientation averaged version of \eqref{eq:T-matrix_general} the isotropic T-matrix, which is given by terms in the diagonal of \eqref{eq:T-matrix_general}, as follows:
\begin{equation}
    \label{eq:diagonal_T-matrix}
    \overline{T}_{n}(\lambda) = - \frac{\mathrm J'_n(kb)}{\mathrm H'_n(kb)} - \frac{\mathrm i z(\lambda)}{\pi [\mathrm H'_n(kb)]^2}.
\end{equation}

\section{Low-frequency behaviour of the resonance factor}
\label{app:expansion}

In this section we show $z(\lambda)$ for a 2D thin walled-split resonator. Other expressions for $z(\lambda)$ for other Helmholtz resonators can be deduced from~\cite{Krynkin_2011, Scat_perforated_cylinders, Split_ring_lipmann_scwinger}  

We use the results from \cite{Atom-Helmholtz}, where $z(\lambda)$ is related to the function $h(\lambda)$ through:
\begin{equation}
    \label{eq:int_geom_factor}
    z(\lambda) = \frac{2 \mathrm i}{\pi (kb)^2 h(\lambda)}.
\end{equation}

For a thin-walled Helmholtz resonator ($b=a$ in \Cref{fig:split-ring_resonator}) $h(\lambda)$ is given by \cite[equation (3.7)]{Atom-Helmholtz}:
\begin{equation}
    \label{eq:h_expression} 
    h(kb, k\ell) =
        2 +
        \frac{4 \mathrm i}{\pi} \left( \gamma_e + \log{\frac{k\ell}{4}} \right) - \frac{1}{2} \sum_{m = -\infty}^{\infty} \frac{\mathrm Q_m^2(kb)}{ \mathrm H'(kb) \mathrm J'(kb)},
\end{equation}
where $\mathrm Q_m(x) = \mathrm J_m(x) \mathrm H_m'(x) + \mathrm J_m'(x) \mathrm H_m(x)$ and $\gamma_e$ is the Euler-Mascheroni constant. For the expression for $h(kb, ka, k\ell)$ for a thick-walled resonator ($a < b$) see \cite[equation (3.7)]{Atom-Helmholtz}, which we use to produce the results in \Cref{fig:three-species-example}.

Performing the asymptotic expansion of $\epsilon = k b \ll 1$ in \eqref{eq:h_expression} we obtain
\begin{equation}
    \label{eq:h_expansion} 
    h(kb, k\ell) =
        2 +
        \frac{4 \mathrm i}{\pi} \left( \gamma_e + \log{\frac{k\ell}{4}} \right) + \frac{2 \mathrm i}{\pi} \frac{1}{\epsilon^2} + \mathcal O \left( \frac{1}{\epsilon} \right),
\end{equation}
where we do not expand the logarithmic term. From the above, and noting that $k \ell < k b$, we can conclude that $|h| \in \mathcal{O}(\epsilon^2)$ for frequencies away from resonance, which implies that $|z| \in \mathcal {O}(1)$. At resonance we have that $h(\lambda)$ is purely real, which occurs when $h(kb, k\ell) = 2$, and then $|z| \in \mathcal O(\epsilon^{-2})$. In other words, $z(\lambda)$ changes its order in $\epsilon$ when passing through resonance, which complicates an asymptotic analysis. 
Luckily, we do not need to perform an asymptotic expansion on $z(\lambda)$ to deduce simple effective formulae for the leading order term.

\bibliographystyle{unsrt}
\bibliography{References.bib}

\end{document}